%% file: hep.tex
\documentclass[12pt,a4paper]{article}
\usepackage{a4wide,epsfig}
\usepackage{axodraw4j}
\usepackage{amsmath}
\usepackage{rotating}
\usepackage{slashed}
\usepackage{alltt}
\usepackage{varwidth}
\usepackage{multirow}

\newcommand{\agt}{\rlap{\lower 3.5 pt \hbox{$\mathchar \sim$}} \raise 1pt
 \hbox {$>$}}
\newcommand{\alt}{\rlap{\lower 3.5 pt \hbox{$\mathchar \sim$}} \raise 1pt
 \hbox {$<$}}

\newcommand{\Li}{\mathop{\mathrm{Li}}\nolimits}
\newcommand{\tol}[1]{\(\overline{\mbox{#1}}\)}

\catcode`@=11
\newcount\@tempcntc
\def\@citex[#1]#2{\if@filesw\immediate\write\@auxout{\string\citation{#2}}\fi
  \@tempcnta\z@\@tempcntb\m@ne\def\@citea{}\@cite{\@for\@citeb:=#2\do
    {\@ifundefined
       {b@\@citeb}{\@citeo\@tempcntb\m@ne\@citea\def\@citea{,}{\bf
?}\@warning
       {Citation `\@citeb' on page \thepage \space undefined}}%
    {\setbox\z@\hbox{\global\@tempcntc0\csname b@\@citeb\endcsname\relax}%
     \ifnum\@tempcntc=\z@ \@citeo\@tempcntb\m@ne
       \@citea\def\@citea{,}\hbox{\csname b@\@citeb\endcsname}%
     \else
      \advance\@tempcntb\@ne
      \ifnum\@tempcntb=\@tempcntc
      \else\advance\@tempcntb\m@ne\@citeo
      \@tempcnta\@tempcntc\@tempcntb\@tempcntc\fi\fi}}\@citeo}{#1}}
\def\@citeo{\ifnum\@tempcnta>\@tempcntb\else\@citea\def\@citea{,}%
  \ifnum\@tempcnta=\@tempcntb\the\@tempcnta\else
   {\advance\@tempcnta\@ne\ifnum\@tempcnta=\@tempcntb \else
\def\@citea{--}\fi
    \advance\@tempcnta\m@ne\the\@tempcnta\@citea\the\@tempcntb}\fi\fi}
\catcode`@=12
\begin{document}

\title{
\vskip-3cm{\baselineskip14pt
\centerline{\normalsize DESY 20--032\hfill ISSN 0418-9833}
\centerline{\normalsize February 2020\hfill}}
\vskip1.5cm
Dipole Subtraction vs.\ Phase Space Slicing in NLO NRQCD Heavy-Quarkonium Production Calculations}

\author{Mathias Butenschoen, Bernd A. Kniehl\\
{\normalsize II. Institut f\"ur Theoretische Physik, Universit\"at Hamburg,}\\
{\normalsize Luruper Chaussee 149, 22761 Hamburg, Germany}
}
\date{\today}

\maketitle

\begin{abstract}
We compare two approaches to evaluate cross sections of heavy-quarkonium production at next-to-leading order in nonrelativistic QCD involving $S$- and $P$-wave Fock states: the customary approach based on phase space slicing and the approach based on dipole subtraction recently elaborated by us.
We find reasonable agreement between the numerical results of the two implementations, but the dipole subtraction implementation outperforms the phase space slicing one both with regard to accuracy and speed.

\medskip

\end{abstract}

\newpage

%%%%%%%%%%%%%%%%%%%%%%%%%%%%%%%%%%%%%%%%%%%%%%%%%%%%%%%%%%%%

\section{Introduction}

The conjectured factorization theorem~\cite{Bodwin:1994jh} of nonrelativistic QCD (NRQCD)~\cite{Caswell:1985ui} is the most frequently used framework for calculations of inclusive heavy-quarkonium production.
It is based on a factorization into perturbative short-distance cross sections for heavy-quark-antiquark pairs in certain Fock states $n$, and nonperturbative long-distance matrix elements (LDMEs).
The numerical values of the latter are extracted from fits to experimental data, and are predicted to scale with certain powers of the relative heavy-quark-antiquark velocity $v$ \cite{Lepage:1992tx}.
For the phenomenologically important quarkonia, the leading and next-to-leading contributions in the $v$ expansion involve $S$- and $P$-wave bound states.
Many calculations of these contributions have been performed at next-to-leading order (NLO) in the strong-coupling constant $\alpha_s$.
These works were almost exclusively done using the two-cutoff phase space slicing scheme as described in Ref.~\cite{Harris:2001sx}.
To our knowledge, the only exception is the work of Ref.~\cite{Campbell:2007ws}, where color-singlet $S$-wave-state production was treated in the massless Catani-Seymour dipole subtraction scheme~\cite{Catani:1996vz}.
In Ref.~\cite{Butenschoen:2019lef}, we have formulated a subtraction scheme covering $S$- and $P$-wave color-singlet and color-octet states for the important example of hadroproduction.
It is based on Ref.~\cite{Catani:1996vz} and its extension to massive quarks by Phaf and Weinzierl \cite{Phaf:2001gc}.
In particular, it takes into account the bound-state structure of the projected amplitudes and introduces new kinds of subtraction terms for the case of $P$-wave-state production.

This paper describes a numerical comparison of our implementations of two-cutoff phase space slicing and dipole subtraction for inclusive quarkonium hadroproduction.
In Section~\ref{sec:Divs}, we briefly review the singularity structure of the encountered real-correction squared amplitudes and their cancellation by other contributions.
We summarize phase space slicing in Section~\ref{sec:slicing} and dipole subtraction in Section~\ref{sec:dipole}, providing many previously unpublished technical details.
In Section~\ref{sec:CompSlicing}, we then numerically compare the two implementations, before summarizing our results in Section~\ref{sec:summary}.

\section{Singular cross section contributions\label{sec:Divs}}

The factorization theorems of QCD and NRQCD imply that the inclusive cross section to produce a heavy-quarkonium state $H$ is given by
\begin{equation}
d\sigma(AB\to H+X)
=\sum_{a,b,X} \sum_n \int dx_a dx_b\, f_{a/A}(x_a)f_{b/B}(x_b)
\langle{\cal O}^{H}[n]\rangle d\hat{\sigma}(a b\to Q\overline{Q}[n]+X), \label{eq:GeneralFactorization}
\end{equation}
with the partonic cross sections
\begin{eqnarray}
d\hat{\sigma}(a b\to Q\overline{Q}[n]+X) &=&
\frac{1}{N_\mathrm{col}(n) N_\mathrm{pol}(n)}\,\frac{1}{2(p_1+p_2)^2}\, d\mathrm{PS} \nonumber \\
&&{}\times \frac{F_\mathrm{sym}(X)}{n_\mathrm{col}(a) n_\mathrm{pol}(a) n_\mathrm{col}(b) n_\mathrm{pol}(b)} \| | a b \to Q\overline{Q}[n]+X\rangle \|^2.
\end{eqnarray}
Here, $f_{a/A}(x_a)$ is the parton distribution function (PDF) describing the probability to find parton $a$ with longitudinal momentum fraction $x_a$ inside hadron $A$.
$\langle{\cal O}^{H}[n]\rangle$ is the LDME of NRQCD associated with the intermediate Fock state $n$, which has $N_\mathrm{col}$ color and $N_\mathrm{pol}$ polarization degrees of freedom.
$p_1$ and $p_2$ are the four-momenta of partons $a$ and $b$, $n_\mathrm{col}$ and $n_\mathrm{pol}$ their color and spin averaging factors.
$d\mathrm{PS}$ is the phase space and $F_\mathrm{sym}$ the symmetry factor associated with the outgoing particles.
$| a b \to Q\overline{Q}[n]+X\rangle$ denotes the matrix element of the partonic subprocess $a b \to Q\overline{Q}[n]+X$, which is calculated by applying spin and color projectors to the usual QCD amplitudes as described in Ref.~\cite{Butenschoen:2019lef}. A summation of spin and color degrees of freedom of the $Q\overline{Q}$ pair and all incoming and outgoing partons is always implicitly understood in the squared amplitudes, but no averaging.
At this point, we deviate from the definition of the bra and ket symbols used in
Refs.~\cite{Catani:1996vz,Phaf:2001gc}.
We denote the momentum of the $Q\overline{Q}$ pair as $p_0$ and set $p_0^2=4m_Q^2$, with $m_Q$ being the heavy-quark mass.
Our real-correction partonic amplitudes have two further light QCD partons, to which we assign momenta $p_3$ and $p_4$.

In the limit where an outgoing gluon with momentum $p_j$ gets soft, the squared production amplitude becomes, for the Fock states considered in our analysis,
\begin{eqnarray}
\| |^1\!S_0^{[1/8]},p_j\;\mathrm{soft}\rangle \|^2 &=& S_1 (^1\!S_0^{[1/8]}; p_j), \nonumber\\
\| |^3\!S_1^{[1/8]},p_j\;\mathrm{soft}\rangle \|^2 &=& S_1 (^3\!S_1^{[1/8]}; p_j), \nonumber\\
\| |^1\!P_1^{[1/8]},p_j\;\mathrm{soft}\rangle \|^2
 &=& S_1 (^1\!P_1^{[1/8]}; p_j) + S_2(^1\!P_1^{[1/8]}, ^1\!S_0^{[1/8]}; p_j) + S_3(^1\!S_0^{[1/8]}; p_j), \nonumber\\
\| |^3\!P_J^{[1/8]},p_j\;\mathrm{soft}\rangle \|^2
 &=& S_1 (^3\!P_J^{[1/8]}; p_j) + S_2(^3\!P_J^{[1/8]}, ^3\!S_1^{[1/8]}; p_j) + S_3(^3\!S_1^{[1/8]}; p_j),
\end{eqnarray}
with
\begin{eqnarray}
S_1(n; p_j)
&=&  g_s^2 \sum_{\substack{i,k=1 \\ i,k\neq j}}^4 \left(-\frac{p_i\cdot p_k}{p_i \cdot p_j \;p_k\cdot p_j} + \frac{p_0\cdot p_i}{p_0\cdot p_j\; p_i\cdot p_j} + \frac{p_0\cdot p_k}{p_0\cdot p_j\; p_k\cdot p_j} - \frac{p_0^2}{(p_0\cdot p_j)^2} \right) \nonumber \\
&&{}\times \langle n, \mathrm{Born} | \mathbf{T}_i \mathbf{T}_k | n, \mathrm{Born} \rangle \label{eq:soft1singular}
\\
S_2(n,m; p_j) &=& 4g_s^2\sum_{\substack{i=1 \\ i\neq j}}^4 \left( \frac{-p_i^\beta}{p_i\cdot p_j\; p_0\cdot p_j} + \frac{p_0\cdot p_i\; p_j^\beta}{p_i\cdot p_j (p_0\cdot p_j)^2} -\frac{p_0^2 p_j^\beta}{(p_0\cdot p_j)^3} \right) \nonumber \\
&&{}\times \epsilon_\beta(m_l) \langle n,\mathrm{Born} | \mathbf{T}_i (\mathbf{T}_Q-\mathbf{T}_{\overline{Q}}) | m,\mathrm{Born}\rangle \label{eq:s2term}
\\
S_3(m; p_j) &=& 4 g_s^2 \left( - \frac{g^{\alpha\beta}}{(p_0\cdot p_j)^2} - \frac{p_0^2 p_j^\alpha p_j^\beta}{(p_0\cdot p_j)^4} \right) \nonumber
\\
&&{}\times \epsilon^\ast_\alpha(m_l)\epsilon_\beta(m_l) \langle m,\mathrm{Born} |  (\mathbf{T}_Q-\mathbf{T}_{\overline{Q}}) (\mathbf{T}_Q-\mathbf{T}_{\overline{Q}}) | m,\mathrm{Born}\rangle \label{eq:s3term},
\end{eqnarray}
where $|m,\mathrm{Born}\rangle$ is the Born amplitude of $Q\overline{Q}[m]$ production without the soft gluon.
$\mathbf{T}_i$ acts on $|m,\mathrm{Born}\rangle$ by inserting at the corresponding place $T_c$ if parton $i$ is an outgoing quark or incoming antiquark, $-T_c$ if parton $i$ is an incoming quark or outgoing antiquark, and $if_{abc}$ if parton $i$ is a gluon, where $c$, $a$, and $b$ are the color indices of the soft,
splitting, and other outgoing gluons, respectively.
$\mathbf{T}_Q$ inserts $T_c$ at the place of the outgoing heavy quark $Q$, $\mathbf{T}_{\overline{Q}}$ inserts $-T_c$ at the place of the outgoing heavy antiquark $\overline{Q}$, with $c$ being the color index of the outgoing gluon attached to the $Q$ or $\overline{Q}$ lines. $\epsilon(m_l)$ is the polarization four-vector of the $Q\overline{Q}[m]$ state with $m_l$ being the quantum number of the $z$ component of its orbital angular momentum.

In the limit where an outgoing light parton with momentum $p_j$ becomes collinear to an incoming light parton with momentum $p_i$, its main contribution stems from Feynman diagrams where parton $i$ splits into $j$ and a parton with momentum $p_{(ij)}=p_i-p_j$ taking away the fraction $x$ of the incoming parton's longitudinal momentum.
The squared matrix element in that limit is given by
\begin{eqnarray}
 \| |p_j \;\mathrm{ini.\;coll.}\; p_i \rangle \|^2 &=& \frac{n_\mathrm{col}(i)
 }{n_\mathrm{col}((ij)) n_\mathrm{pol}((ij)) } \, \frac{g_s^2}{x(p_i\cdot p_j)} \,
 \langle \mathrm{Born} | \hat{P}_{i,(ij)}(x,p_\perp) | \mathrm{Born} \rangle \nonumber \\
 &&{}\times
 \begin{cases}
  \delta_{ss^\prime} & \mbox{if $i$ is a quark or antiquark} \\
  \epsilon_\mu^\ast(p_i) \epsilon_\nu(p_i) & \mbox {if $i$ is a gluon}
 \end{cases}, \label{eq:IniColl}
\end{eqnarray}
where $\hat{P}_{i,(ij)}(x,p_{\perp })$ are the spin-dependent Altarelli-Parisi splitting functions  as given in Eqs. (39)--(42) of Ref.~\cite{Butenschoen:2019lef} with $p_{\perp }$ being the residual transverse component of $p_{(ij)}$. The $\hat{P}_{i,(ij)}(x,p_{\perp })$ functions depend on the spin indices $s$ and $s'$ or the polarization indices $\mu$ and $\nu$
of parton $i$. The squared amplitude in the limit where the outgoing partons 3 and 4 are collinear is given by those Feynman diagrams where a final-state parton with momentum $p_{(34)}=p_3+p_4$ splits into the outgoing partons 3 and 4, and reads
\begin{eqnarray}
 \| |p_3 \;\mathrm{final\;coll.}\; p_4 \rangle \|^2 &=& \frac{g_s^2}{p_3\cdot p_4}
 \langle \mathrm{Born} | \hat{P}_{(34),3}(z,p_\perp) | \mathrm{Born} \rangle ,
 \label{eq:FinColl}
\end{eqnarray}
where $p_3=z p_{(34)} + {\cal O}(p_{\perp })$, $p_{\perp }$ is the residual transverse component of $p_3$, and the open spin or polarization indices within $\hat{P}_{(34),3}(z,p_{\perp })$ match the corresponding open indices of parton (34) in the Born amplitude.

The phase space integrations in $D=4-2\epsilon$ dimensions yield $\frac{1}{\epsilon}$ and $\frac{1}{\epsilon^2}$ poles, which are canceled by similar poles in the virtual corrections, by the {\em mass factorization counterterms}, and by the {\em operator renormalization counterterms}:
A part of the initial-state collinear singularities is absorbed into the PDFs according to the $\overline{\mathrm{MS}}$ prescription, thereby leading to mass factorization counterterms,
\begin{eqnarray}
\lefteqn{d\hat\sigma_\mathrm{MFC}(a+b\to Q\overline{Q}[n]+X) = \left[ \sum_{(ij)} \int dx P_{a,(ij)}^+(x) d\hat\sigma_\mathrm{Born}((ij)+b\to Q\overline{Q}[n]+X) \right.} \nonumber \\
 &&{}+ \left. \sum_{(ij)} \int dx P_{b,(ij)}^+(x) d\hat\sigma_\mathrm{Born}(a+(ij)\to Q\overline{Q}[n]+X)\right] \frac{g_s^2}{8\pi^2} \left( \frac{4\pi\mu_r^2}{\mu_f^2}e^{-\gamma_E}\right)^\epsilon \frac{1}{\epsilon},\label{eq:MFC}
\end{eqnarray}
where $\mu_r$ is the renormalization scale, $\mu_f$ is the QCD factorization scale, and $P_{a,(ij)}^+(x)$ are the regularized Altarelli-Parisi splitting functions as listed in Ref.~\cite{Butenschoen:2019lef}.
The singularities of the $S_3$ part of the soft singularities are canceled by NLO corrections to LDMEs, where ultraviolet singularities are removed by $\overline{\mathrm{MS}}$ renormalization. These {\em operator renormalization} contributions are, for the Fock states relevant to our analysis, given by
\begin{eqnarray}
d\sigma_{^3S_1^{[1]}+^3S_1^{[8]}\, \mathrm{op.ren.}}
&=&\sum_{\substack{a,b,X\\2\to 2}}\sum_{c=1,8}\sum_J \int dx_a dx_b\, f_{a/A}(x_a)f_{b/B}(x_b)
\frac{\langle{\cal O}^{H}[^3P_J^{[c]}]\rangle}{N_\mathrm{col}(^3P_J^{[c]}) N_\mathrm{pol}(^3P_J^{[c]})} \nonumber \\
&&{}\times \frac{1}{2(p_1+p_2)^2}\, d\mathrm{PS}_2
\frac{F_\mathrm{sym}(X)}{n_\mathrm{col}(a) n_\mathrm{pol}(a) n_\mathrm{col}(b) n_\mathrm{pol}(b)} \| | ^3P_J^{[c]},\,\mathrm{op.ren.}\rangle \|^2,\qquad
\label{eq:OpRenTransformed3PJOne}
\end{eqnarray}
with
\begin{eqnarray}
 \| | ^3P_J^{[c]},\mathrm{op.ren.}\rangle \|^2 &=& \frac{g_s^2}{12\pi^2 m_Q^2} \left( \frac{4\pi\mu_r^2}{\mu_\Lambda^2} e^{-\gamma_E} \right)^\epsilon g^{\alpha\beta}\left(-\frac{1}{\epsilon}\right) \nonumber \\
&&{}\times \epsilon_\alpha^\ast(m_l)\epsilon_\beta(m_l) \langle ^3S_1^{[c]}, \mathrm{Born} | (\mathbf{T}_Q-\mathbf{T}_{\overline{Q}}) (\mathbf{T}_Q-\mathbf{T}_{\overline{Q}}) | ^3S_1^{[c]}, \mathrm{Born} \rangle,\qquad
 \label{eq:OpRenTransformed3PJTwo}
\end{eqnarray}
and
\begin{eqnarray}
d\sigma_{^1S_0^{[1]}+^1S_0^{[8]}\, \mathrm{op.ren.}}
&=&\sum_{\substack{a,b,X\\2\to 2}}\sum_{c=1,8} \int dx_a dx_b\, f_{a/A}(x_a)f_{b/B}(x_b)
\frac{\langle{\cal O}^{H}[^1P_1^{[c]}]\rangle}{N_\mathrm{col}(^1P_1^{[c]}) N_\mathrm{pol}(^1P_1^{[c]})} \nonumber \\
&&{}\times \frac{1}{2(p_1+p_2)^2}\, d\mathrm{PS}_2
\frac{F_\mathrm{sym}(X)}{n_\mathrm{col}(a) n_\mathrm{pol}(a) n_\mathrm{col}(b) n_\mathrm{pol}(b)} \| | ^1P_1^{[c]},\,\mathrm{op.ren.}\rangle \|^2,\qquad
\label{eq:OpRenTransformed1P1One}
\end{eqnarray}
with
\begin{eqnarray}
 \| | ^1P_1^{[c]},\mathrm{op.ren.}\rangle \|^2 &=& \frac{g_s^2}{12\pi^2 m_Q^2} \left( \frac{4\pi\mu_r^2}{\mu_\Lambda^2} e^{-\gamma_E} \right)^\epsilon g^{\alpha\beta} \left(-\frac{1}{\epsilon}\right) \nonumber \\
&&{}\times \epsilon_\alpha^\ast(m_l)\epsilon_\beta(m_l) \langle ^1S_0^{[c]}, \mathrm{Born} | (\mathbf{T}_Q-\mathbf{T}_{\overline{Q}}) (\mathbf{T}_Q-\mathbf{T}_{\overline{Q}}) | ^1S_0^{[c]}, \mathrm{Born} \rangle,\qquad
\label{eq:OpRenTransformed1P1Two}
\end{eqnarray}
where $\mu_\Lambda$ is the NRQCD factorization scale.

\section{Phase space slicing implementation\label{sec:slicing}}

Our implementation of phase space slicing follows the lines of Ref.~\cite{Harris:2001sx}.
Here, the real-correction phase space is split into three regions by introducing two cut-off parameters, $\delta_s$ and $\delta_c$:
The soft region, where $p_3$ or $p_4$ is soft, the hard-collinear region, where $p_3$ and $p_4$ are hard and $p_3$ or $p_4$ is collinear to another massless parton, and the hard-noncollinear region.
The condition of $p_j$ being soft is defined by $\delta_s>2E_j/\sqrt{s}$ with $E_{j}$ the energy component of $p_j$ in the center-of-mass frame of $p_1$ and $p_2$, and the condition of $p_i$ being collinear to $p_j$ by $\delta_c > |2p_i\cdot p_j|/\sqrt{s}$ with $s=(p_1+p_2)^2$.
Since the hard-noncollinear region is free of singularities, the phase space integration is done there numerically, while, in the other two regions, the phase space integrations are done analytically in $D=4-2\epsilon$ dimensions.
This is possible because not only the squared matrix elements factorize as described above, but also the phase space elements factorize as $d\mathrm{PS}_3=d\mathrm{PS}_2 d\mathrm{PS}_{p_j}$, where $p_j$ is soft, and $d\mathrm{PS}_3=d\mathrm{PS}_2 d\mathrm{PS}_{p_i\parallel p_j}$, where $p_i$ is collinear to $p_j$. Here, $d\mathrm{PS}_3$ is the phase-space factor of the process $p_1+p_2\to p_0+p_3+p_4$, $d\mathrm{PS}_2$ is the phase-space factor of the Born process corresponding to the respective soft or collinear limit, and
\begin{equation}
 d\mathrm{PS}_{p_j}=\frac{d^{D-1}p_j}{2(2\pi)^{D-1}E_j}, \qquad d\mathrm{PS}_{p_i\parallel p_j}=
 \begin{cases}
 \frac{d^{D-1}p_j}{2(2\pi)^{D-1}E_j} & \mbox{if } i=1,2 \mbox{ and } j=3,4
\\
 \frac{d^{D-1}p_j}{2(2\pi)^{D-1}}\, \frac{E_{(ij)}}{E_i E_j} & \mbox{if } i \mbox{ and } j=3,4
 \end{cases}
.
\end{equation}
The dependencies of all contributions on $\delta_s$ and $\delta_c$ cancel in the sum, as long as $\delta_s$ and $\delta_c$ are chosen small enough.

\subsection{Hard-collinear part}

Integrating Eq.~(\ref{eq:IniColl}) analytically over the hard-collinear phase space part and adding the corresponding contribution of the mass factorization counterterm in Eq.~(\ref{eq:MFC}), we obtain in the limit $\delta_s\to 0$
\begin{eqnarray}
\left[\int_{p_i\parallel p_j} d\mathrm{PS}_{p_i\parallel p_j} \| |p_i \mathrm{\,ini.\,coll.\,} p_j \rangle \|^2\right]_{+\mathrm{MFC}} &=& \frac{n_\mathrm{pol}(i) n_\mathrm{col}(i)}{n_\mathrm{pol}((ij)) n_\mathrm{col}((ij))} \int_{x_\mathrm{min}}^{1-\delta_s \delta_{j,g}} \frac{dx}{x}  \| | \mathrm{Born} \rangle \|^2 \nonumber \\
&&\times \Bigg[ \delta_{j,g} \delta(1-x) F_{\mathrm{in}, i\to (ij),j} \nonumber \\
&&\quad+ \frac{g_s^2}{8\pi^2}  \left( P_{i,(ij)}(x) \ln \frac{(1-x)\delta_c s}{\mu_f^2} - P^\prime_{i,(ij)}(x) \right) \Bigg],
\nonumber\\
&&
\end{eqnarray}
where the Born amplitude $| \mathrm{Born} \rangle$ is defined with an incoming momentum $p_{(ij)}=x p_i$ instead of $p_i$.
$\delta_{j,g}$ is 1 if particle $j$ is a gluon, otherwise 0.
$F_{\mathrm{in}, i\to (ij),j}$ are given by
\begin{eqnarray}
F_{\mathrm{in},g\to gg} &=& \frac{g_s^2}{8\pi^2} C_\epsilon \left( \frac{1}{\epsilon} +  \ln \frac{m_Q^2}{\mu_f^2} \right) \left( 2 C_A \ln \delta_s + \frac{11}{6}C_A - \frac{n_f}{3}\right), \nonumber\\
F_{\mathrm{in},q\to qg} &=& \frac{g_s^2}{8\pi^2} C_\epsilon C_F \left( \frac{1}{\epsilon} + \ln \frac{m_Q^2}{\mu_f^2} \right) \left( 2 \ln \delta_s+\frac{3}{2}\right),
\end{eqnarray}
with $C_A=3$, $C_F=4/3$, $n_f$  is the number of light, active quark flavors, and $C_\epsilon=(4\pi\mu_r^2e^{-\gamma_E}/m_Q^2)^\epsilon$.
Furthermore, $P_{i,(ij)}$ and $P^\prime_{i,(ij)}$ are the ${\cal O}(\epsilon^0)$ and ${\cal O}(\epsilon)$ parts of the spin-averaged splitting functions, namely
\begin{eqnarray}
 P_{qq}(x) &=& C_F \frac{1+x^2}{1-x}, \nonumber\\
 P^\prime_{qq}(x) &=& -C_F(1-x), \nonumber\\
 P_{qg}(x) &=& C_F \frac{1+(1-x)^2}{x}, \nonumber\\
 P^\prime_{qg}(x) &=& -C_F x, \nonumber\\
 P_{gg}(x) &=& 2C_A \left( \frac{x}{1-x} + \frac{1-x}{x} + x(1-x)\right), \nonumber\\
 P^\prime_{gg}(x) &=& 0, \nonumber\\
 P_{gq}(x) &=& \frac{1}{2}\left( x^2 + (1-x)^2 \right), \nonumber\\
 P^\prime_{gq}(x) &=& -x(1-x).
\end{eqnarray}

Integrating Eq.~(\ref{eq:FinColl}) analytically over the hard-collinear phase space part, we obtain in the limit $\delta_s\to 0$
\begin{eqnarray}
\int_{p_3\parallel p_4} d\mathrm{PS}_{p_3\parallel p_4} \| |p_3 \mathrm{\,fin.\,coll.\,} p_4 \rangle \|^2 = \| | \mathrm{Born} \rangle \|^2 F_{\mathrm{fi}, (34)\to 3,4},
\end{eqnarray}
with
\begin{eqnarray}
F_{\mathrm{fi}, g\to g g} &=&  \frac{g_s^2 C_A}{8\pi^2} C_\epsilon \left[ \left( \frac{1}{\epsilon} - \ln \frac{\delta_cs}{m_Q^2}\right) \left( \frac{11}{6} + 2\ln \frac{s\delta_s}{s-4m_Q^2}\right) - \ln^2 \frac{\delta_ss}{s-4m_Q^2} + \frac{67}{18} - \frac{\pi^2}{3} \right], \nonumber\\
F_{\mathrm{fi}, g\to q\overline{q}} &=& \frac{g_s^2}{8\pi^2} \frac{n_f}{3} C_\epsilon \left[ -\frac{1}{\epsilon} + \ln \frac{\delta_c s}{m_Q^2} - \frac{5}{3} \right], \nonumber\\
F_{\mathrm{fi}, q\to q g} &=& \frac{g_s^2 C_F}{8\pi^2} C_\epsilon \left[ \left( \frac{1}{\epsilon} - \ln \frac{\delta_c s}{m_Q^2} \right) \left( \frac{3}{2}+2 \ln\frac{\delta_s s}{s-4m_Q^2}\right) -\ln^2 \frac{\delta_ss}{s-4m_Q^2} + \frac{7}{2}-\frac{\pi^2}{3} \right].\qquad
\end{eqnarray}

\subsection{Soft part: $S_1$ terms}

The integral of the $S_1$ terms in Eq.~(\ref{eq:soft1singular}) over the soft phase space region can be written as
\begin{eqnarray}
 \int_{p_j\;\mathrm{soft}} d\mathrm{PS}_{p_j} S_1(n,p_j) = g_s^2\left( -C_{00} I_{00} - \sum_{\substack{i,k=0\\\{i,j,k\}\,\mathrm{distinct}}}^4 C_{ik} I_{ik} \right),
\end{eqnarray}
with
\begin{eqnarray}
 C_{ik} = \langle \mathrm{Born} | \mathbf{T}_i \mathbf{T}_k | \mathrm{Born} \rangle, \qquad I_{ik} = \int_{p_j\,\mathrm{soft}} d\mathrm{PS}_{p_j} \frac{ p_i\cdot p_k }{p_i\cdot p_j\;p_k\cdot p_j},
\end{eqnarray}
where we define $\mathbf{T}_0 = \mathbf{T}_Q + \mathbf{T}_{\overline{Q}}$ and use $\mathbf{T}_i=-\sum_{\substack{k=0\\k\neq i,j}}^4 \mathbf{T}_k$, with $p_j$ being the soft momentum.
Evaluating the integrals $I_{ik}$ following Ref.~\cite{Harris:2001sx}, we obtain
\begin{eqnarray}
 I_{0,0} &=& \frac{1}{8\pi^2} C_\epsilon \left[ -\frac{1}{\epsilon} - \frac{s+4m_Q^2}{s-4m_Q^2} \ln \frac{s}{4m_Q^2} + \ln \frac{\delta_s^2 s}{m_Q^2} \right],
\nonumber \\
 I_{0,k=1\,\mathrm{or}\,2} &=& \frac{1}{16\pi^2} C_\epsilon \left[\frac{1}{\epsilon^2} - \frac{1}{\epsilon} \ln \frac{\psi_k^2 \delta_s^2}{4m_Q^4} + \ln^2\frac{-\psi_k}{4m_Q^2} - \frac{1}{2}\ln^2\frac{s}{m_Q^2} + \frac{1}{2}\ln^2\frac{\delta_s^2s}{m_Q^2} \right.\nonumber \\
 &&{}+ \left. \ln\frac{\psi_k^2}{4m_Q^2s} \ln \frac{\delta_s^2s}{m_Q^2} + 2 \Li_2\left( \frac{\psi_k+4m_Q^2}{4m_Q^2}\right) - 2\Li_2\left( \frac{-\xi_k}{\psi_k}\right) - \frac{\pi^2}{4} \right],
\nonumber \\
I_{0,k=3\,\mathrm{or}\,4} &=& \frac{1}{16\pi^2} C_\epsilon \left[ \frac{1}{\epsilon^2} - \frac{1}{\epsilon} \ln \frac{\delta_s^2 s^2}{4m_Q^4} + \frac{1}{2}\ln^2 \frac{\delta_s^2s^2}{4m_Q^4} + 2 \Li_2\left(\frac{4m_Q^2-s}{4m_Q^2}\right) - \frac{\pi^2}{4} \right],
\nonumber \\
I_{i=1\,\mathrm{or}\,2,k=1\,\mathrm{or}\,2} &=& \frac{1}{8\pi^2} C_\epsilon \left[ \frac{1}{\epsilon^2} - \frac{1}{\epsilon} \ln \frac{\delta_s^2s}{m_Q^2} + \frac{1}{2}\ln^2\frac{\delta_s^2s}{m_Q^2} - \frac{\pi^2}{4} \right],
\nonumber \\
I_{i=1\,\mathrm{or}\,2,k=3\,\mathrm{or}\,4} &=& \frac{1}{8\pi^2} C_\epsilon \left[ \frac{1}{\epsilon^2} - \frac{1}{\epsilon} \ln \frac{-\delta_s^2 s \xi_i}{(s-4m_Q^2)m_Q^2} + \frac{1}{2}\ln^2 \frac{-\delta_s^2 s \xi_i}{(s-4m_Q^2)m_Q^2} \right.\nonumber \\
&&{}+ \left. \Li_2\left(\frac{-\psi_i-4m_Q^2}{s-4m_Q^2}\right) - \frac{\pi^2}{4} \right],
\end{eqnarray}
with $\psi_i= -2p_0\cdot p_i$, $\xi_i = -2p_{(34)}\cdot p_i$, and $p_{(34)}=p_3+p_4-p_j$.

\subsection{Soft part: $S_2$ terms}

The integral of the $S_2$ terms in Eq.~(\ref{eq:s2term}) over the soft phase space region is
\begin{eqnarray}
  \lefteqn{\int_{p_j\,\mathrm{soft}} d\mathrm{PS}_{p_j} S_2(n,m,p_j) = 4g_s^2 \epsilon_\beta(m_l)\sum_{\substack{i=1\\ i\neq j}}^4\langle n, \mathrm{Born} | \mathbf{T}_i (\mathbf{T}_Q-\mathbf{T}_{\overline{Q}}) |m, \mathrm{Born}\rangle} \nonumber \\
 &&{}\times\underbrace{\int_{p_j\,\mathrm{soft}} d\mathrm{PS}_{p_j} \left( - \frac{p_i^\beta}{p_i\cdot p_j\; p_0\cdot p_j} + \frac{p_0\cdot p_i \;p_j^\beta}{p_i\cdot p_j(p_0\cdot p_j)^2} - \frac{p_0^2 p_j^\beta}{(p_0\cdot p_j)^3} \right)}_{=\Omega_{i,j}^\beta}. \label{eq:Soft2Integral}
\end{eqnarray}
To evaluate the phase space integrals involving $p_j^\beta$, we use the tensor decomposition
\begin{eqnarray}
 \int_{p_j\,\mathrm{soft}} d\mathrm{PS}_{p_j} \left( \frac{p_0\cdot p_i}{p_i\cdot p_j (p_0\cdot p_j)^2} - \frac{p_0^2}{(p_0\cdot p_j)^3}\right) p_j^\beta =
 \begin{cases}
   A_1 p_i^\beta + A_2 p_{(34)}^\beta + A_3 p_0^\beta & \mbox{for }i=1,2 \\
   A_4 p_{(34)}^\beta + A_5 p_0^\beta  & \mbox{for }i=3,4
 \end{cases},
\nonumber\\
\label{eq:TensRedSoft2}
\end{eqnarray}
which leads to the expressions
\begin{eqnarray}
 \Omega_{i=1\,\mathrm{or}\,2,j}^\beta &=& p_i^\beta \int_{p_j\,\mathrm{soft}} d\mathrm{PS}_{p_j} \left[ \frac{\left(s-4m_Q^2\right)\left(\psi_i s+4m_Q^2(s+3\xi_i)\right)}{2\left(\psi_i+4m_Q^2\right)s\xi_i}\,\frac{1}{(p_0\cdot p_j)^2} \right. \nonumber \\
 &&{}- \frac{\psi_i s - 4 m_Q^2 (\psi_i+2\xi_i)}{2 \left(\psi_i+4 m_Q^2\right)s}\, \frac{1}{p_i\cdot p_j\; p_0\cdot p_j} + \frac{4 m_Q^2\left(\psi_i s + 4m_Q^2(s-\xi_i)\right)}{\left(\psi_i+4m_Q^2\right)s\xi_i}\,\frac{p_3\cdot p_4}{(p_0\cdot p_j)^3} \nonumber \\
 &&{}+\left. \frac{4 m_Q^2 \left(s-4 m_Q^2\right)^2}{\left(\psi_i+4m_Q^2\right)s\xi_i}\, \frac{p_i\cdot p_j}{\left(p_0\cdot p_j\right)^3} + \frac{\psi_i \left(\psi_i s - 4 m_Q^2(\psi_i+2\xi_i)\right)}{2\left(\psi_i+4m_Q^2\right)s\xi_i}\, \frac{p_3\cdot p_4}{(p_0\cdot p_j)^2 \;p_i\cdot p_j} \right]
 \nonumber \\
 &&{}+ p_{(34)}^\beta \int_{p_j\,\mathrm{soft}} d\mathrm{PS}_{p_j} \left[ \frac{4m_Q^2\psi_i^2}{\left(\psi_i+4m_Q^2\right)s\xi_i}\, \frac{p_3\cdot p_4}{(p_0\cdot p_j)^3} - \frac{\psi_i^2}{2\left(\psi_i+4m_Q^2\right)s}\,\frac{1}{p_0\cdot p_j \;p_i\cdot p_j}  \right. \nonumber \\
 &&{}+ \frac{4\left(\psi_ism_Q^2 - 4 m_Q^4(\psi_i+2\xi_i)\right)}{\left(\psi_i+4m_Q^2\right)s\xi_i}\,\frac{p_i\cdot p_j}{\left(p_0\cdot p_j\right)^3} +  \frac{\psi_i^3}{2\left(\psi_i+4m_Q^2\right)s\xi_i}\,\frac{p_3\cdot p_4}{(p_0\cdot p_j)^2 \;p_i\cdot p_j}  \nonumber \\
 &&{}+ \left.   \frac{\psi_i\left(\psi_is - 4m_Q^2(\psi_i+4\xi_i)\right)}{2\left(\psi_i+4m_Q^2\right)s\xi_i}\,\frac{1}{(p_0\cdot p_j)^2} \right] + \left(p_0^\beta\,\mathrm{terms}\right),
\nonumber \\
\Omega_{i=3\,\mathrm{or}\,4,j}^\beta &=& p_{(34)}^\beta \int_{p_j\,\mathrm{soft}} d\mathrm{PS}_{p_j} \left( - \frac{2 p_0^2}{(p_0\cdot p_j)^2 \;p_0\cdot p_i} +\frac{\left(p_0^2\right)^2\; p_i\cdot p_j}{(p_0\cdot p_j)^3 (p_0\cdot p_i)^2} \right) + \left(p_0^\beta\,\mathrm{terms}\right),\qquad
\end{eqnarray}
where the $p_0^\beta$ terms vanish upon contraction with $\epsilon_\beta(m_l)$ in Eq.~(\ref{eq:Soft2Integral}).
As for $\Omega_{i=1\,\mathrm{or}\,2,j}$, the angular integrals needed to evaluate $\int_{p_j\,\mathrm{soft}} d\mathrm{PS}_{p_j} p_i\cdot p_j/(p_0\cdot p_j)^3$ and $\int_{p_j\,\mathrm{soft}} d\mathrm{PS}_{p_j} p_3\cdot p_4/( (p_0\cdot p_j)^2(p_i\cdot p_j))$ are not listed in
Ref.~\cite{Harris:2001sx} or the references cited therein.
We obtain these by relating the phase space integrals to cut virtual-correction loop integrals and evaluating the latter by means of the integration-by-parts technique \cite{Tkachov:1981wb}.
The final results are
\begin{eqnarray}
\Omega_{i=1\,\mathrm{or}\,2,j}^\beta &=&
 - \frac{C_\epsilon p_i^\beta}{4\pi^2\psi_i} \left[ \frac{1}{\epsilon} - \frac{\psi_i^2+\psi_i \xi_i + 4 m_Q^2 (\psi_i+2\xi_i)}{\xi_i\left(\psi_i+4m_Q^2\right)} \ln \frac{-\psi_i}{4m_Q^2} -\frac{s}{\xi_i}\ln\frac{s}{4m_Q^2}  - \ln \frac{\delta_s^2s}{m_Q^2} \right] \nonumber \\
 &&{}- \frac{p_{(34)}^\beta}{4\pi^2\xi_i} \Bigg[ \frac{\left(s^2-16m_Q^4\right)\xi_i}{\left(s-4m_Q^2\right)^3} + \frac{\psi_i}{\psi_i+4m_Q^2} \ln \frac{-\psi_i}{4m_Q^2}\nonumber \\
 &&{}+
\frac{16m_Q^4\psi_i + s^2\left(\psi_i+8m_Q^2\right)}{\left(s-4m_Q^2\right)^3} \ln \frac{s}{4m_Q^2}  \Bigg] + \left(p_0^\beta\,\mathrm{terms}\right),
\nonumber \\
\Omega_{i=3\,\mathrm{or}\,4,j}^\beta &=& \frac{C_\epsilon p_{(34)}^\beta}{4\pi^2(s-4m_Q^2)}  \left[ \frac{1}{\epsilon} + 1 - \frac{2s}{s-4m_Q^2} + \frac{2s^2}{\left(s-4m_Q^2\right)^2}\ln\frac{s}{4m_Q^2} - \ln \frac{\delta_s^2 s^2}{4m_Q^4} \right]  \nonumber \\
 &&{}+ \left(p_0^\beta\,\mathrm{terms}\right).
\end{eqnarray}

\subsection{Soft part: $S_3$ terms}

The integral of the $S_3$ terms in Eq.~(\ref{eq:s3term}) over the soft phase space region is
\begin{eqnarray}
\int_{p_j\,\mathrm{soft}} d\mathrm{PS}_{p_j} S_3(m,p_j)&=& 4 g_s^2 \epsilon_\alpha^\ast(m_l) \epsilon_\beta(m_l) \langle m, \mathrm{Born} | (\mathbf{T}_Q-\mathbf{T}_{\overline{Q}})(\mathbf{T}_Q-\mathbf{T}_{\overline{Q}}) |m,\mathrm{Born}\rangle \nonumber \\
&&{}\times \underbrace{\int_{p_j\,\mathrm{soft}} d\mathrm{PS}_{p_j} \left( -\frac{g^{\alpha\beta}}{(p_0\cdot p_j)^2} - \frac{p_0^2\; p_j^\alpha p_j^\beta}{(p_0\cdot p_j)^4}\right)}_{=\Omega_j^{\alpha\beta}}.
\label{eq:soft3integral}
\end{eqnarray}
To evaluate the integral involving $p_j^\alpha p_j^\beta$, we use the tensor decomposition
\begin{eqnarray}
 \int_{p_j\,\mathrm{soft}} d\mathrm{PS}_{p_j} \frac{p_j^\alpha p_j^\beta}{(p_0\cdot p_j)^4} = A_6 g^{\alpha\beta} + A_7 p_{(34)}^\alpha p_{(34)}^\beta + A_8 p_0^\alpha p_0^\beta + A_9 \left(p_0^\alpha p_{(34)}^\beta + p_{(34)}^\alpha p_0^\beta\right), \label{eq:TensRedSoft3}
\end{eqnarray}
resulting in
\begin{eqnarray}
\Omega_j^{\alpha\beta} &=& \frac{g^{\alpha\beta}}{\epsilon-1} \int_{p_j\,\mathrm{soft}} d\mathrm{PS}_{p_j} \left[ \frac{32 m_Q^4}{\left(s-4m_Q^2\right)^2}\,\frac{(p_3\cdot p_4)^2}{(p_0\cdot p_j)^4}   -   \frac{8m_Q^2}{s-4m_Q^2}\,\frac{p_3\cdot p_4}{(p_0\cdot p_j)^3} -\frac{\epsilon-1}{(p_0\cdot p_j)^2} \right] \nonumber \\
&&{}+ \frac{16 p_{(34)}^\alpha p_{(34)}^\beta}{\epsilon-1} \int_{p_j\,\mathrm{soft}} d\mathrm{PS}_{p_j} \left[ \frac{32(3-2\epsilon) m_Q^6}{\left(s-4m_Q^2\right)^4}\,\frac{(p_3\cdot p_4)^2}{(p_0\cdot p_j)^4} - \frac{(\epsilon-1)m_Q^2}{\left(s-4m_Q^2\right)^2}\,\frac{1}{(p_0\cdot p_j)^2} \right. \nonumber \\
&&{}- \left. \frac{8(3-2\epsilon) m_Q^4}{\left(s-4m_Q^2\right)^3}\,\frac{p_3\cdot p_4}{\left(p_0\cdot p_j\right)^3}  \right] + \left(p_0\,\mathrm{terms}\right),
\end{eqnarray}
where the $p_0^\alpha$ and $p_0^\beta$ terms vanish upon contraction with $\epsilon_\beta(m_l)$ in Eq.~(\ref{eq:soft3integral}).
Evaluating the integrals analytically and adding the corresponding operator renormalization counterterm contribution of Eqs.~(\ref{eq:OpRenTransformed3PJOne}) or~(\ref{eq:OpRenTransformed1P1One}), we arrive at the finite expression
\begin{eqnarray}
 \left( \Omega_j^{\alpha\beta} \right)_{+\mathrm{op.\,ren.}} &=& \frac{g^{\alpha\beta}}{48m_Q^2\pi^2}  \left[ - \frac{\left(s+4m_Q^2\right)^2}{2\left(s-4m_Q^2\right)^2} - \frac{1}{2}\ln \frac{4\delta_s^4s}{m_Q^2} - \ln\frac{m_Q^2}{\mu_\Lambda^2}  \right. \nonumber \\
  &&{}+\left. \left( \frac{2s^3}{\left(s-4m_Q^2\right)^3} - \frac{3s^2}{\left(s-4m_Q^2\right)^2} + \frac{6m_Q^2}{s-4m_Q^2} + \frac{3s}{2\left(s-4m_Q^2\right)}\right) \ln \frac{s}{4m_Q^2}  \right]   \nonumber \\
  &&{}+\frac{p_{(34)}^\alpha p_{(34)}^\beta}{\pi^2} \Bigg[ -\frac{2s^2}{\left(s-4m_Q^2\right)^4} + \frac{2s}{\left(s-4m_Q^2\right)^3} - \frac{1}{6\left(s-4m_Q^2\right)^2} \nonumber \\
  &&{}+ \frac{2s^3 - 3s^2\left(s-4m_Q^2\right) + s\left(s-4m_Q^2\right)^2}{\left(s-4m_Q^2\right)^5} \ln \frac{s}{4m_Q^2} \Bigg]  + \left(p_0\,\mathrm{terms}\right).
\end{eqnarray}

\subsection{A remark on the tensor decomposition}

We note that the tensor decompositions of Eqs.~(\ref{eq:TensRedSoft2}) and (\ref{eq:TensRedSoft3}) involve momentum $p_{(34)}$.
A potential pitfall is that $p_{(34)}$ does not appear in the integrand of
Eq.~(\ref{eq:TensRedSoft3}), nor does it in Eq.~(\ref{eq:TensRedSoft2}) in the cases $i=1$ or 2, and one might na\"{\i}vely think that $d\mathrm{PS}_{p_j}$ only involves momentum $p_j$, of which $p_{(34)}$ is independent in the limit of $p_j$ being soft.
Thus, one might be led to assume that the structures with $p_{(34)}$ are not necessary in the above-mentioned cases.
However, the dependence on $p_{(34)}$ does enter in a more subtle way, via the phase space constraint of $p_j$ being soft, which implies
$\delta_s>2E_j/\sqrt{s}=(s-2p_0\cdot p_{(34)}-4m_Q^2)/s$.
Alternatively, to understand the necessity of all structures involving $p_{(34)}$, one can convince oneself that the corresponding coefficients $A_2$, $A_4$, $A_7$, and $A_9$ in Eqs.~(\ref{eq:TensRedSoft2}) and (\ref{eq:TensRedSoft3}) are indeed nonzero.
To this end, one applies the projectors
\begin{eqnarray}
\Pi_2^\beta &=& \frac{\left(\psi_i s + 4 m_Q^2 (s-\xi_i)\right)
p_i^\beta + \psi_i^2 p_{(34)}^\beta - \psi_i \xi_i
p_0^\beta}{s\xi_i\left(s-4m_Q^2+\xi_i\right)},
\nonumber\\
\Pi_4^\beta &=& \frac{-16m_Q^2 p_{(34)}^\beta + 2\left(s-4m_Q^2\right)
  p_0^\beta}{\left(s-4m_Q^2\right)^2},
\nonumber\\
\Pi_7^{\alpha\beta} &=& \frac{256 m_Q^4 (D-1) p_{(34)}^\alpha
p_{(34)}^\beta}{(D-2)\left(s-4m_Q^2\right)^4} - \frac{32 m_Q^2 (D-1)
\left(p_{(34)}^\alpha p_0^\beta + p_0^\alpha
p_{(34)}^\beta\right)}{(D-2)\left(s-4m_Q^2\right)^3}
\nonumber\\
&&{}+ \frac{16m_Q^2 g^{\alpha\beta} + 4 (D-2) p_0^\alpha
  p_0^\beta}{(D-2)\left(s-4m_Q^2\right)^2},
\nonumber\\
\Pi_9^{\alpha\beta}&=&
-\frac{2g^{\alpha\beta}}{(D-2)\left(s-4m_Q^2\right)} - \frac{32 m_Q^2
(D-1) p_{(34)}^\alpha p_{(34)}^\beta}{(D-2)\left(s-4m_Q^2\right)^3} +
\frac{2D\left(p_{(34)}^\alpha p_0^\beta + p_0^\alpha
p_{(34)}^\beta\right)}{(D-2)\left(s-4m_Q^2\right)^2},
\end{eqnarray}
and inserts the results of the soft phase space integrals.
By the same token, using a tensor decomposition with the $p_{(34)}$ structures omitted then leads to incorrect results for $\Omega_{i=1\,\mathrm{or}\,2,j}^\beta$ and $\Omega_j^{\alpha\beta}$.
Incidentally, this neither spoils the infrared finiteness nor the dependence on $\delta_s$, which would have served as a crucial check otherwise.
So, this mistake is easily made and more easily overlooked. 
As already mentioned in Ref.~\cite{Butenschoen:2019npa},
Ref.~\cite{Butenschoen:2012px} is affected by this mistake.
The numerical effects of this are discernible in Fig.~2 of Ref.~\cite{Butenschoen:2012px}, where the dashed curves, indicating the ${}^3\!P_J^{[8]}$
contributions, are subject to visible deviations.
Fortunately, the effects on the physical results in Fig.~1 of
Ref.~\cite{Butenschoen:2012px} are insignificant, being of the order of the
numerical uncertainty.
We believe that this easy-to-miss mistake has also creeped into other authors'
calculations.
In fact, Eqs.~(5) and (6) of Ref.~\cite{Jia:2014jfa} only hold if the
incorrect tensor decomposition is applied.
Furthermore, purposely including this mistake, we are able to approximately
reproduce the results shown in Fig.~3 of Ref.~\cite{Gong:2012ug} and Fig.~2 of
Ref.~\cite{Feng:2018ukp}, while our correct evaluations significantly differ
from these results.

\section{Dipole subtraction implementation\label{sec:dipole}}

\subsection{Summary of dipole subtraction formalism}

Our implementation of dipole subtraction, which is based on Refs.~\cite{Catani:1996vz,Phaf:2001gc} is explained in detail in Ref.~\cite{Butenschoen:2019lef}.
For the reader's convenience, we recall the main formulas here.
We calculate the NLO corrections to our partonic cross sections as
\begin{eqnarray}
 \int d\hat{\sigma} &=& \int d\mathrm{PS}_3 \left[ \frac{d\hat{\sigma}_\mathrm{real}}{d\mathrm{PS}_3} \theta(p_T-p_{T,\mathrm{min}})-\frac{d\hat{\sigma}_\mathrm{subtr}}{d\mathrm{PS}_3} \theta(\tilde{p}_T-p_{T,\mathrm{min}}) \right]
 \nonumber \\
 &&{}+\int d\mathrm{PS}_2 \left[ \frac{d\hat{\sigma}_\mathrm{virtual}+d\hat{\sigma}_\mathrm{MFC}+d\hat{\sigma}_\mathrm{op.\,ren.}}{d\mathrm{PS}_2} \theta(p_T-p_{T,\mathrm{min}}) \right. \nonumber \\
 &&{}+\left. [dx] \theta(\tilde{p}_T-p_{T,\mathrm{min}}) \int d\mathrm{PS}_\mathrm{dipole} \frac{d\hat{\sigma}_\mathrm{subtr}}{d\mathrm{PS}_3} \right],
 \label{eq:dipolesubwithcuts}
\end{eqnarray}
where $d\mathrm{PS}_2$ and $d\mathrm{PS}_3$ are the two- and three-particle phase space factors.
The latter factorize in some way as $d\mathrm{PS}_2d\mathrm{PS}_\mathrm{dipole}$ or $d\mathrm{PS}_2 dxd\mathrm{PS}_\mathrm{dipole}$, where $dx$ matches its counterpart in Eq.~(\ref{eq:MFC}).
$d\hat{\sigma}_\mathrm{real}$, $d\hat{\sigma}_\mathrm{virtual}$, $d\hat{\sigma}_\mathrm{MFC}$, and $d\hat{\sigma}_\mathrm{op.\,ren.}$ are the real-correction contributions, the virtual-correction contributions, the mass factorization counterterms, and operator renormalization counterterms, respectively.
The {\em subtraction term} $d\hat{\sigma}_\mathrm{subtr}$ is given by
\begin{eqnarray}
 \frac{d\hat{\sigma}_\mathrm{subtr}(a+b\to Q\overline{Q}[n]+X)}{d\mathrm{PS}_3} &=& \frac{1}{N_\mathrm{col}(n) N_\mathrm{pol}(n)}\,\frac{1}{2(p_1+p_2)^2} \nonumber \\
 &&{}\times \frac{F_\mathrm{sym}(X)}{n_\mathrm{col}(a) n_\mathrm{pol}(a) n_\mathrm{col}(b) n_\mathrm{pol}(b)} \, \| | abn, \mathrm{subtr}\rangle \|^2,
 \label{eq:dipolesmoredetailedOne}
\end{eqnarray}
with
\begin{eqnarray}
 \| | abn, \mathrm{subtr} \rangle \|^2 &=& \sum_{j=3}^4 \sum_{i=1}^2 \sum_{\substack{k=0\\k\neq i,j}}^4
 \frac{n_\mathrm{col}(i)}{n_\mathrm{col}((ij))}
  \frac{-1}{2p_i\cdot p_j} \frac{1}{x} \langle n, \mathrm{Born} | V^{\mathrm{ini,}S_1}_{ij,k}\frac{\mathbf{T}_{(ij)} \mathbf{T}_k}{\mathbf{T}_{(ij)}^2} | n, \mathrm{Born} \rangle 
\nonumber \\
&&{}+ \sum_{j=3}^4 \sum_{\substack{i=0 \\ i\neq 1,2,j}}^3 \sum_{\substack{k=0\\k\neq i,j}}^4 \frac{-1}{2p_i\cdot p_j} \langle n, \mathrm{Born} | V^{\mathrm{fin,}S_1}_{ij,k}\frac{\mathbf{T}_{(ij)} \mathbf{T}_k}{\mathbf{T}_{(ij)}^2} | n, \mathrm{Born} \rangle
\begin{cases}
\frac{1}{x} & \mathrm{if}\,k=1,2 \\ 1 & \mathrm{if}\,k\neq1,2
\end{cases}
\nonumber \\
&&{}+ \sum_{j=3}^4 \sum_{\substack{i=1 \\ i\neq j}}^4  V_{S_2,ij}^{\beta} \epsilon_\beta(m_l) \langle n,\mathrm{Born} | \mathbf{T}_{(ij)} (\mathbf{T}_Q-\mathbf{T}_{\overline{Q}}) | m(n),\mathrm{Born}\rangle
\nonumber \\
&&{}+ \sum_{j=3}^4 V_{S_3,j}^{\alpha\beta} \epsilon^\ast_\alpha(m_l)\epsilon_\beta(m_l) \langle m(n),\mathrm{Born} |  (\mathbf{T}_Q-\mathbf{T}_{\overline{Q}}) (\mathbf{T}_Q-\mathbf{T}_{\overline{Q}}) | m(n),\mathrm{Born}\rangle.
\nonumber\\
&&\label{eq:dipolesmoredetailed}
\end{eqnarray}
This term is defined in terms of $2\to 3$ kinematic variables in the same way as $d\hat{\sigma}_\mathrm{real}$ and constructed so that it matches $d\hat{\sigma}_\mathrm{real}$ in all singular limits:
We call each term in the sum a {\em dipole}.
The dipoles in the first and second lines of Eq.~(\ref{eq:dipolesmoredetailed}) reproduce $d\hat{\sigma}_\mathrm{real}$ in the initial- and final-state collinear limits as well as the $S_1$ part of the soft limits, while the dipoles in the last two lines reproduce the $S_2$ and $S_3$ parts of the soft limits.
As for the Born amplitudes in Eq.~(\ref{eq:dipolesmoredetailed}), partons $i$ and $j$ are replaced by $(ij)$, which is a gluon, light quark, or the $Q\overline{Q}$ pair, according to the soft or collinear limits to be approximated.
The contribution in Eq.~(\ref{eq:dipolesmoredetailed}) is set to zero if
there is no collinear or soft singularity in the considered limit.
In the dipoles for the $S_2$ and $S_3$ terms, $m({^3P}_J^{[1/8]})={^3S}_1^{[1/8]}$ and $m({^1P}_1^{[1/8]})={^1S}_0^{[1/8]}$.
Since the Born amplitudes are defined in terms of $2\to 2$ kinematic variables, we need for each dipole a {\em mapping} of the $2\to 3$ process momenta to $2\to 2$ kinematics momenta $\tilde{p}_i$.
These momentum mappings are constructed in such a way that the Born amplitudes in Eq.~(\ref{eq:dipolesmoredetailed}) equal their counterparts in the factorization formulae of the respective collinear and soft limits.
Furthermore, we define
\begin{equation}
 \tilde{p}_T^2 = \frac{(4m_Q^2-\tilde{t})(\tilde{s}+\tilde{t})}{\tilde{s}}-4m_Q^2,
 \qquad
 \tilde{y} = \ln \frac{\tilde{s} + \tilde{t}}{\tilde{x}_2 \sqrt{S\left(\tilde{p}_T^2+4m_Q^2\right)}},
\end{equation}
with
\begin{equation}
 \tilde{s}=(\tilde{p}_1+\tilde{p}_2)^2, \qquad \tilde{t}=(\tilde{p}_0-\tilde{p}_1)^2, \qquad S=(p_A+p_B)^2, \qquad \tilde{p}_2=\tilde{x}_2 p_B,
\end{equation}
where $p_A$ and $p_B$ are the momenta of the incoming hadrons.

The idea of the subtraction formalism is that the term in the first square bracket of Eq.~(\ref{eq:dipolesubwithcuts}) is devoid of singularities and can, therefore, be integrated over numerically in four dimensions.
On the other hand, the various $V$ terms in Eq.~(\ref{eq:dipolesmoredetailed}) are defined only in terms of those kinematic variables that we use to parametrize $d\mathrm{PS}_\mathrm{dipole}$, and are sufficiently simple to be integrated analytically over $d\mathrm{PS}_\mathrm{dipole}$.
The resulting poles in $\epsilon$ then render the second square bracket in Eq.~(\ref{eq:dipolesmoredetailed}) finite and ready for numerical integration over $d\mathrm{PS}_2$ and $dx$.

In practice, we need to produce predictions involving experimental cuts on the transverse momentum $p_T$ and the rapidity $y$ of the heavy quarkonium, see for example the low-$p_T$ cut in Eq.~(\ref{eq:dipolesubwithcuts}).
The implementation of these phase space cuts is, however, unproblematic, too, since, in the first square bracket of Eq.~(\ref{eq:dipolesubwithcuts}), $p_T$ and $\tilde{p}_T$ coincide in the singular limits and, in the second square bracket, the $\theta$ function stands outside the analytic $d\mathrm{PS}_\mathrm{dipole}$ integration. Similarly, $y$ and $\tilde{y}$ coincide in the singular limits, too.

Table~\ref{tab:DipoleOverview} indicates where one can find the analytic expressions for the various $V$ terms, their counterparts upon integration over the
respective phase space $d\mathrm{PS}_\mathrm{dipole}$, and their respective momentum mappings.
As for the $V$ terms of the Catani-Seymour~\cite{Catani:1996vz} and Phaf-Weinzierl~\cite{Phaf:2001gc} papers, $V^{\mathrm{ini,}S_1}_{ij,k}$ equals $V^{ij}_k$ or $V^{ij,k}$ and $V^{\mathrm{fin,}S_1}_{ij,k}$ equals $V_{ij,k}$ or $V_{ij}^k$ in their notations.
Indices $s$ and $s^\prime$ or $\mu$ and $\nu$ within $V_{ij,k}$ are to be contracted with the open polarization indices of particle $(ij)$ in the corresponding Born amplitude.
Figure~\ref{fig:listofdipoles} summarizes all dipole terms according to their corresponding singular limits in a graphical form.

\begin{table}
\begin{center}
\begin{tabular}{l|cc|l|l}
& $p_i$ & $p_k$ & Definition and Integration & Applied Mapping
\\
\hline
 $V^{\mathrm{ini,}S_1}_{ij,k}$ &$p_1$ or $p_2$ & $p_0$ & PW, section 6.1 & MapPW6($p_i, p_j$)\\
 $V^{\mathrm{ini,}S_1}_{ij,k}$ &$p_1$ or $p_2$ & $p_1$ or $p_2$ & CS, section 5.6 ($n=p_3+p_4$)& MapCS($p_i$) \\
 $V^{\mathrm{ini,}S_1}_{ij,k}$ &$p_1$ or $p_2$ & $p_3$ or $p_4$ & CS, section 5.3 & MapCS($p_i$) \\
 $V^{\mathrm{fin,}S_1}_{ij,k}$ &$p_0$ & $p_1$ or $p_2$ & PW, section 6.2 & MapPW6($p_k, p_j$) \\
 $V^{\mathrm{fin,}S_1}_{ij,k}$&$p_0$ & $p_3$ or $p_4$ & PW, section 5.2 & MapPW5.2($p_j$) \\
 $V^{\mathrm{fin,}S_1}_{ij,k}$&$p_3$ or $p_4$ & $p_0$ & PW, section 5.1 & MapPW5.1($p_i$)\\
 $V^{\mathrm{fin,}S_1}_{ij,k}$&$p_3$ or $p_4$ & $p_1$ or $p_2$ & CS, section 5.2 & MapCS($p_k$) \\
 $V_{S_2,ij}$ & $p_1$ or $p_2$ && BK, (75) and section 4.5.1 & MapPW6($p_i,p_j$) \\
 $V_{S_2,ij}$ & $p_3$ or $p_4$ && BK, (75) and section 4.5.2 & MapPW5.2($p_j$) \\
 $V_{S_3,j}$ && & BK, (76) and section 4.5.3 & MapPW5.2($p_j$)
\end{tabular}
\end{center}
\caption{\label{tab:DipoleOverview}%
List of occurring $V$ terms with given momentum assignments; of where their definitions and analytic expressions upon integration over the dipole phase spaces may be found in the Catani-Seymour (CS) \cite{Catani:1996vz}, Phaf-Weinzierl (PW) \cite{Phaf:2001gc}, and Butenschoen-Kniehl (BK) \cite{Butenschoen:2019lef} papers; and of momentum mappings, according to the naming scheme of Ref.~\cite{Butenschoen:2019lef}, to be applied to the numerical integrations of the respective dipole terms over $d\mathrm{PS}_3$.}
\end{table}

\input{TableOfDipoles}

\subsection{Organization in terms of computer codes}

In this section, we briefly describe how we implement the dipole subtraction procedure in our computer codes, emphasizing those parts which differ from our
implementation of phase space slicing.
All necessary Born diagrams are created with {\tt FeynArts} and then treated by a Mathematica script which inserts the color operators $\mathbf{T}$ in the various combinations needed and applies the color-singlet and color-octet projectors to evaluate all color factors in the squared amplitudes.
These color factors, together with the yet unsquared amplitudes, are then passed to two FORM scripts, {\tt CalcDipoles} and {\tt CalcDipolesInteg}.
For these two routines, we have prepared an input card which encodes the information of Figure~\ref{fig:listofdipoles}.

{\tt CalcDipoles} generates the FORTRAN routines encoding the dipole terms.
This is done by squaring the Born amplitudes, written in terms of the $\tilde{p}_i$ momenta and with the respective color insertions, and multiplying them by the necessary factors, in particular $V_{ij,k}$, taking into account the spin correlations in the case of splitting gluons.
Then, the respective momentum mappings are implemented, the resulting expressions are simplified, and the FORTRAN routine {\tt AMP2\_Dipoles} is generated, which takes as arguments the number of the dipole and the partonic $2\to 3$ kinematic variables.

Similarly, {\tt CalcDipolesInteg} squares the Born amplitudes with the respective color insertions, and then multiplies the finite parts of the $V$ terms integrated over the dipole phase space $d\mathrm{PS}_{\mathrm{dipole}}$.
At this point, the mass factorization and operator renormalization counterterms are included, as described in Ref.~\cite{Butenschoen:2019lef}.
The integrated dipoles have the general form $[h(x)]_+f(x)+g(x)$, where $h(x)$ is singular in the limit $x\to1$.
The generated FORTRAN function {\tt AMP2\_DipolesInteg} takes as arguments the number of the dipole, the $2\to 2$ kinematics variables, and the value of $x$.
A second function, {\tt AMP2\_Dipoles\-IntegSubtr}, which only contains the terms $h(x)f(1)$, is generated as well.

Together with the FORTRAN functions for the virtual- and the real-correction squared amplitudes, we then have the ingredients for the numerical phase space integrations in the main FORTRAN code. The $\theta$ functions constraining the $2\to 3$ particle phase space, such as $\theta(p_T-p_{T,\mathrm{min}})$ in Eq.~(\ref{eq:dipolesubwithcuts}), have to be implemented for each dipole individually with the respective momentum mapping.
The $x$ integrations over the plus distributions are thereby explicitly implemented as
 \begin{equation}
  \int_{x_\mathrm{min}}^1 dx \big\{ [h(x)]_+ f(x) + g(x) \big\} = \int_0^1dx \Big\{ \theta(x-x_\mathrm{min}) \underbrace{\big[h(x)f(x)+g(x)\big]}_{\tt AMP2\_DipolesInteg} - \underbrace{h(x)f(1)}_{\substack{\tt AMP2\_Dipoles\\ \tt IntegSubtr}} \Big\},
 \end{equation}
so that the singularities of the $h(x)$ terms for $x\to1$ cancel numerically in the course of the integration.

\subsection{Numerical tests of the integrated dipole terms\label{sec:NumDipoleTest}}

In Ref.~\cite{Butenschoen:2019lef}, we have already described one numerical test of our dipole subtraction implementation, namely we have shown that our expressions for the dipole terms agree with the real-correction contributions in the corresponding singular limits.
Here, we describe a further internal test.
This time, we test the expressions of the integrated dipole terms.
We do this by evaluating the phase space integrals,
\begin{eqnarray}
 {\cal I}_i = \int d\mathrm{PS}_3 \theta(p_T-p_{T,\mathrm{min}}) \left( \| | abn, \mbox{subtr} \rangle_{\mathrm{dipole}\,i} \|^2 \right)_\mathrm{+MFC\,+op.\,ren.}, \label{eq:Iidef}
\end{eqnarray}
of specific dipole terms plus the corresponding mass factorization and operator renormalization counterterms in two different ways and comparing the results.
In the first mode of evaluation, we use the results of the expressions implemented in {\tt CalcDipolesInteg} and integrate them numerically over $d\mathrm{PS}_2$ or $d\mathrm{PS}_2 dx$.
In the second mode of evaluation, we separate the three-particle phase space as in the phase space slicing implementation according to the slicing parameters $\delta_s$ and $\delta_c$.
For the contributions from the soft and collinear regions, we take the respective analytic limits of the dipole terms, integrate them analytically over the corresponding phase space, $d\mathrm{PS}_{p_j\,\mathrm{soft}}$ or $d\mathrm{PS}_{i\parallel j}$, add the corresponding mass factorization and operator renormalization counterterms, and then do the integrations over $d\mathrm{PS}_2$ or $d\mathrm{PS}_2dx$ numerically.
For the contribution from the hard-noncollinear region, we integrate the expressions for the dipole terms as encoded in {\tt CalcDipoles} numerically over $d\mathrm{PS}_3$.
Both contributions are then combined to yield the final results of the second
mode of evaluation.
We perform these tests with groups of one, two, or three dipoles in order to simplify the analytic integrations of the soft limits in the second version.
We have successfully tested all the dipoles in this way.
Typical examples are presented in Table~\ref{tab:dipoleintegtestp}.

\begin{table}
\begin{center}
\begin{varwidth}{\linewidth}
\begin{scriptsize}
\begin{alltt}
DipoleIntegtest gg2cCg for State 3PJ8.
Dipole  8 on.
Dipole 13 on.
 
 Result Analytic:    20193.01298070771 \(\pm\) 2.996135930530795     
 
 Result Slicing:
  0.000001      -1.4614E+05 \(\pm\) 1.0517E+01     1.6171E+05 \(\pm\) 8.8167E+02     1.5567E+04 \(\pm\) 8.8173E+02
  0.000010      -1.0832E+05 \(\pm\) 8.3362E+00     1.2746E+05 \(\pm\) 3.4805E+02     1.9130E+04 \(\pm\) 3.4815E+02
  0.000100      -7.4543E+04 \(\pm\) 5.6847E+00     9.4589E+04 \(\pm\) 2.0637E+02     2.0047E+04 \(\pm\) 2.0644E+02
  0.001000      -4.4854E+04 \(\pm\) 3.6687E+00     6.4916E+04 \(\pm\) 8.2790E+01     2.0063E+04 \(\pm\) 8.2871E+01
  0.010000      -1.9291E+04 \(\pm\) 1.7919E+00     3.9340E+04 \(\pm\) 3.2753E+01     2.0049E+04 \(\pm\) 3.2802E+01

DipoleIntegtest gd2cCd for State 3P21.
Dipole 243 on.
 
 Result Analytic:   -170.1669990971855 \(\pm\) 1.5578494469294161E-004
 
 Result Slicing:
  0.000001       4.5608E+02 \(\pm\) 4.0023E-04    -6.2616E+02 \(\pm\) 2.3290E-01    -1.7008E+02 \(\pm\) 2.3290E-01
  0.000010       3.6877E+02 \(\pm\) 3.2966E-04    -5.3970E+02 \(\pm\) 2.0804E-01    -1.7093E+02 \(\pm\) 2.0804E-01
  0.000100       2.8332E+02 \(\pm\) 2.7631E-04    -4.5313E+02 \(\pm\) 1.4900E-01    -1.6981E+02 \(\pm\) 1.4900E-01
  0.001000       1.9636E+02 \(\pm\) 1.9533E-04    -3.6631E+02 \(\pm\) 1.1489E-01    -1.6995E+02 \(\pm\) 1.1489E-01
  0.010000       1.0978E+02 \(\pm\) 1.0843E-04    -2.7811E+02 \(\pm\) 7.8774E-02    -1.6833E+02 \(\pm\) 7.8774E-02
\end{alltt}
\end{scriptsize}
\end{varwidth}
\end{center}
\caption{\label{tab:dipoleintegtestp}%
Numerical tests of integrated dipole subtraction terms as described in Section~\ref{sec:NumDipoleTest}.
The finite parts of $I_i$ in Eq.~(\ref{eq:Iidef}) (in GeV$^{-5}$) are evaluated for $g+g\to c\bar{c}[{}^3\!P_J^{[8]}]+g$ with dipoles 8 and 13 and for $g+d\to c\bar{c}[{}^3\!P_J^{[8]}]+d$ with dipole 243 using the implementations of dipole subtraction and phase space slicing, for $n_f=3$, $\alpha_s=1/(4\pi)$, $\mu_f=0.5~\text{GeV}$, $m_Q=0.2~\text{GeV}$,  $(p_1+p_2)^2=100~\text{GeV}^2$, $p_{T,\text{min}}=2~\text{GeV}$, $\delta_c=\delta$, and $\delta_s=\delta/1000$ with variable value of $\delta$.
For a given value of $\delta$ (first column), the soft and collinear parts (second column), the hard-noncollinear parts (third column), and their sum (fourth column) are listed.
The quoted errors are the numerical-integration uncertainties.
}
\end{table}

\section{Comparison of phase space slicing and dipole subtraction methods\label{sec:CompSlicing}}

In Tables~\ref{tab:dipoleslicingcomp}--\ref{tab:dipoleslicingcomp23P21}, we compare our dipole subtraction and phase space slicing implementations.
We calculate at NLO total cross sections, including the Born contributions, of
inclusive  charmonium production in proton-antiproton collisions at typical
center-of-mass-energies for selected bins of transverse momentum and rapidity.
As in Refs.~\cite{Butenschoen:2009zy,Butenschoen:2012px,Butenschoen:2019npa},
we treat the first $n_f=3$ quark flavors as massless,
take the heavy-quark mass, defined in the on-shell renormalization scheme, to
be $m_Q=1.5~\text{GeV}$,
use set CTEQ6M \cite{Pumplin:2002vw} of proton PDFs,
evaluate $\alpha_s=\alpha_s^{(4)}(\mu_r)$ at two loops with asymptotic scale
parameter $\Lambda_\mathrm{QCD}^{(4)}=326~\text{MeV}$ \cite{Pumplin:2002vw}, and
choose the renormalization and factorizations scales to be
$\mu_r=\mu_f=\sqrt{p_T^2+4m_Q^2}$ and $\mu_\Lambda=m_Q$.
For definiteness, we set
$\langle {\cal O}^{H}({}^3\!S_1^{[8]})\rangle=1~\text{GeV}^3$ and
$\langle {\cal O}^{H}({}^3\!P_0^{[1]})\rangle=1~\text{GeV}^5$.
In the phase space slicing implementation, we choose the cut-off parameters to
be $\delta_s=\delta$ and $\delta_c=\delta/1000$, vary $\delta$ from $10^{-5}$ to $10^{-2}$, and take the evaluation with $\delta=10^{-3}$ as default to be
compared with the results obtained using dipole subtraction.

From Tables~\ref{tab:dipoleslicingcomp}--\ref{tab:dipoleslicingcomp23P21}, we observe that the results obtained in the selected bins using the two implementations numerically agree at the level of about 10\%, in line with the uncertainty inherent in the application of the phase space slicing method.
Besides being more accurate, the dipole subtraction implementation is also typically much faster than the phase space slicing implementation.
The reason for that is that, in the phase space slicing implementation, there is usually a much stronger numerical cancellation between the contributions from
the analytic and numerical integrations than in the dipole subtraction implementation, necessitating a higher relative accuracy in the numerical integrations.
However, this advantage is to some extent compensated by the fact that the $\theta$ functions in the first term of Eq.~(\ref{eq:dipolesubwithcuts}) cut out very different phase space regions and so worsen the convergence of the numerical Monte-Carlo integrations in the dipole subtraction implementation.
Nevertheless, we observe that our dipole implementation achieves a final accuracy of 1\% typically 2 to 6 times faster than the phase space slicing implementation.

\begin{table}
\begin{center}
\begin{varwidth}{\linewidth}
%\begin{scriptsize}
\begin{tiny}
\begin{alltt}
p+\tol{p} --> c\tol{c}[3S18] + X:  Sqrt[s] = 100 GeV,  2 GeV < p_T < 3 GeV,  -0.5 < y < 0.5:
================================================================================

Dipole implementation:   1.8694E+03 \(\pm\) 1.8683E+00

Slicing implementation:
  0.000010    -4.2106E+04 \(\pm\) 1.3991E+01     4.3677E+04 \(\pm\) 1.4321E+01     1.5710E+03 \(\pm\) 2.0021E+01
  0.000032    -3.5355E+04 \(\pm\) 1.0905E+01     3.7067E+04 \(\pm\) 9.5207E+00     1.7120E+03 \(\pm\) 1.4477E+01
  0.000100    -2.9207E+04 \(\pm\) 8.8019E+00     3.0987E+04 \(\pm\) 6.1790E+00     1.7800E+03 \(\pm\) 1.0754E+01
  0.000320    -2.3394E+04 \(\pm\) 6.1821E+00     2.5209E+04 \(\pm\) 3.9720E+00     1.8149E+03 \(\pm\) 7.3482E+00
  0.001000    -1.8171E+04 \(\pm\) 4.6479E+00     1.9989E+04 \(\pm\) 2.5723E+00     1.8184E+03 \(\pm\) 5.3122E+00
  0.003200    -1.3329E+04 \(\pm\) 3.3122E+00     1.5139E+04 \(\pm\) 1.6327E+00     1.8103E+03 \(\pm\) 3.6927E+00
  0.010000    -9.0788E+03 \(\pm\) 2.1645E+00     1.0867E+04 \(\pm\) 1.0867E+00     1.7882E+03 \(\pm\) 2.4220E+00

Relative difference using default slicing parameter: -2.7%
Factor of time the dipole implementation is faster: 9.6

p+\tol{p} --> c\tol{c}[3S18] + X:  Sqrt[s] = 100 GeV,  6 GeV < p_T < 7 GeV,  -0.5 < y < 0.5:
================================================================================

Dipole implementation:   2.6433E+01 \(\pm\) 2.5363E-02

Slicing implementation:
  0.000010    -5.9185E+02 \(\pm\) 1.9436E-01     6.1327E+02 \(\pm\) 3.7003E-01     2.1420E+01 \(\pm\) 4.1797E-01
  0.000032    -4.9661E+02 \(\pm\) 1.4844E-01     5.2090E+02 \(\pm\) 2.4433E-01     2.4290E+01 \(\pm\) 2.8589E-01
  0.000100    -4.0982E+02 \(\pm\) 1.1312E-01     4.3508E+02 \(\pm\) 1.0923E-01     2.5260E+01 \(\pm\) 1.5725E-01
  0.000320    -3.2764E+02 \(\pm\) 8.2613E-02     3.5340E+02 \(\pm\) 1.6469E-01     2.5761E+01 \(\pm\) 1.8425E-01
  0.001000    -2.5363E+02 \(\pm\) 5.8880E-02     2.7997E+02 \(\pm\) 2.3740E-01     2.6338E+01 \(\pm\) 2.4459E-01
  0.003200    -1.8482E+02 \(\pm\) 4.0917E-02     2.1062E+02 \(\pm\) 7.1737E-02     2.5795E+01 \(\pm\) 8.2586E-02
  0.010000    -1.2431E+02 \(\pm\) 2.6148E-02     1.4989E+02 \(\pm\) 1.9901E-01     2.5578E+01 \(\pm\) 2.0072E-01

Relative difference using default slicing parameter: -0.4%
Factor of time the dipole implementation is faster: 6.4

p+\tol{p} --> c\tol{c}[3S18] + X:  Sqrt[s] = 1960 GeV,  7 GeV < p_T < 8 GeV,  -0.6 < y < 0.6:
=================================================================================

Dipole implementation:   2.3196E+03 \(\pm\) 2.3170E+00

Slicing implementation:
  0.000010    -6.6859E+04 \(\pm\) 2.0831E+01     6.8617E+04 \(\pm\) 1.7129E+02     1.7580E+03 \(\pm\) 1.7255E+02
  0.000032    -5.6557E+04 \(\pm\) 1.5699E+01     5.8597E+04 \(\pm\) 3.8898E+01     2.0400E+03 \(\pm\) 4.1946E+01
  0.000100    -4.7142E+04 \(\pm\) 1.2016E+01     4.9345E+04 \(\pm\) 4.0183E+01     2.2030E+03 \(\pm\) 4.1941E+01
  0.000320    -3.8205E+04 \(\pm\) 9.0543E+00     4.0421E+04 \(\pm\) 1.0403E+01     2.2164E+03 \(\pm\) 1.3791E+01
  0.001000    -3.0131E+04 \(\pm\) 6.7318E+00     3.2404E+04 \(\pm\) 2.0048E+01     2.2727E+03 \(\pm\) 2.1148E+01
  0.003200    -2.2592E+04 \(\pm\) 4.6217E+00     2.4810E+04 \(\pm\) 3.9682E+00     2.2180E+03 \(\pm\) 6.0915E+00
  0.010000    -1.5909E+04 \(\pm\) 3.2149E+00     1.8126E+04 \(\pm\) 7.9005E+00     2.2171E+03 \(\pm\) 8.5296E+00

Relative difference using default slicing parameter: -2.0%
Factor of time the dipole implementation is faster: 3.7
\end{alltt}
%\end{scriptsize}
\end{tiny}
\end{varwidth}
\end{center}
\caption{\label{tab:dipoleslicingcomp}%
Numerical comparisons of our implementations of dipole subtraction and phase space slicing as described in Section~\ref{sec:CompSlicing},
for $\sigma(p\bar{p}\to c\bar{c}[{}^3\!S_1^{[8]}]+X)$ (in nb) with
$\sqrt{S}=100~\text{GeV}$, $2~\text{GeV}<p_T<3~\text{GeV}$, and $-0.5<y<0.5$;
$\sqrt{S}=100~\text{GeV}$, $6~\text{GeV}<p_T<7~\text{GeV}$, and $-0.5<y<0.5$;
and
$\sqrt{S}=1.96~\text{TeV}$, $7~\text{GeV}<p_T<8~\text{GeV}$, and $-0.6<y<0.6$.
For a given value of $\delta$ (first column), the hard-collinear contribution
of the real corrections (third column), the residual part including the soft and collinear contributions (second column), and their sum (fourth column) are listed.
The quoted errors are the numerical-integration uncertainties.
For the default value $\delta=10^{-3}$, the relative deviations of the phase
space slicing results from the dipole subtraction ones and the factors by which
the dipole subtraction implementation is faster than the space slicing one in
achieving a numerical accuracy of 1\% are indicated.}
\end{table}

\begin{table}
\begin{center}
\begin{varwidth}{\linewidth}
\begin{scriptsize}
\begin{alltt}
p+\tol{p} --> c\tol{c}[3S18] + X:  Sqrt[s] = 7000 GeV,  6 GeV < p_T < 8 GeV,  2.5 < y < 4:
==============================================================================

Dipole implementation:   1.5209E+04 \(\pm\) 1.5039E+01

Slicing implementation:
  0.000010    -4.4765E+05 \(\pm\) 1.3365E+02     4.5846E+05 \(\pm\) 1.6483E+02     1.0810E+04 \(\pm\) 2.1221E+02
  0.000032    -3.7798E+05 \(\pm\) 1.0265E+02     3.9071E+05 \(\pm\) 7.0978E+01     1.2732E+04 \(\pm\) 1.2480E+02
  0.000100    -3.1423E+05 \(\pm\) 8.1206E+01     3.2768E+05 \(\pm\) 7.9244E+01     1.3452E+04 \(\pm\) 1.1346E+02
  0.000320    -2.5395E+05 \(\pm\) 6.2502E+01     2.6789E+05 \(\pm\) 3.1169E+01     1.3943E+04 \(\pm\) 6.9842E+01
  0.001000    -1.9945E+05 \(\pm\) 4.4830E+01     2.1363E+05 \(\pm\) 8.1179E+01     1.4182E+04 \(\pm\) 9.2735E+01
  0.003200    -1.4864E+05 \(\pm\) 3.1923E+01     1.6281E+05 \(\pm\) 1.6280E+01     1.4167E+04 \(\pm\) 3.5834E+01
  0.010000    -1.0368E+05 \(\pm\) 2.1351E+01     1.1778E+05 \(\pm\) 1.8833E+01     1.4097E+04 \(\pm\) 2.8470E+01

Relative difference using default slicing parameter: -6.8%
Factor of time the dipole implementation is faster: 4.4

p+\tol{p} --> c\tol{c}[3S18] + X:  Sqrt[s] = 7000 GeV,  20 GeV < p_T < 22 GeV,  -0.5 < y < 0.5:
===================================================================================

Dipole implementation:   1.6314E+02 \(\pm\) 1.8166E-01

Slicing implementation:
  0.000010    -4.7483E+03 \(\pm\) 1.4145E+00     4.8660E+03 \(\pm\) 1.6862E+00     1.1770E+02 \(\pm\) 2.2009E+00
  0.000032    -4.0204E+03 \(\pm\) 1.0596E+00     4.1628E+03 \(\pm\) 1.0825E+00     1.4245E+02 \(\pm\) 1.5148E+00
  0.000100    -3.3545E+03 \(\pm\) 8.1121E-01     3.5065E+03 \(\pm\) 7.1555E-01     1.5199E+02 \(\pm\) 1.0817E+00
  0.000320    -2.7209E+03 \(\pm\) 6.2684E-01     2.8762E+03 \(\pm\) 4.7362E-01     1.5528E+02 \(\pm\) 7.8565E-01
  0.001000    -2.1458E+03 \(\pm\) 4.6500E-01     2.3026E+03 \(\pm\) 3.0469E-01     1.5681E+02 \(\pm\) 5.5593E-01
  0.003200    -1.6074E+03 \(\pm\) 3.0795E-01     1.7633E+03 \(\pm\) 1.9829E-01     1.5590E+02 \(\pm\) 3.6627E-01
  0.010000    -1.1275E+03 \(\pm\) 2.1164E-01     1.2834E+03 \(\pm\) 1.2832E-01     1.5587E+02 \(\pm\) 2.4750E-01

Relative difference using default slicing parameter: -3.9%
Factor of time the dipole implementation is faster: 1.4

p+\tol{p} --> c\tol{c}[3S18] + X:  Sqrt[s] = 14000 GeV,  5 GeV < p_T < 8 GeV,  2 < y < 4:
=============================================================================

Dipole implementation:   8.9783E+04 \(\pm\) 8.9783E+01

Slicing implementation:
  0.000010    -2.7420E+06 \(\pm\) 8.3086E+02     2.8014E+06 \(\pm\) 1.0012E+03     5.9400E+04 \(\pm\) 1.3010E+03
  0.000032    -2.3177E+06 \(\pm\) 5.9967E+02     2.3901E+06 \(\pm\) 7.6892E+02     7.2450E+04 \(\pm\) 9.7511E+02
  0.000100    -1.9297E+06 \(\pm\) 4.9280E+02     2.0083E+06 \(\pm\) 4.4693E+02     7.8650E+04 \(\pm\) 6.6528E+02
  0.000320    -1.5625E+06 \(\pm\) 3.8080E+02     1.6439E+06 \(\pm\) 2.9593E+02     8.1360E+04 \(\pm\) 4.8227E+02
  0.001000    -1.2304E+06 \(\pm\) 2.8138E+02     1.3132E+06 \(\pm\) 1.9530E+02     8.2850E+04 \(\pm\) 3.4252E+02
  0.003200    -9.2029E+05 \(\pm\) 2.0434E+02     1.0037E+06 \(\pm\) 1.2350E+02     8.3410E+04 \(\pm\) 2.3876E+02
  0.010000    -6.4571E+05 \(\pm\) 1.3564E+02     7.2853E+05 \(\pm\) 7.7187E+01     8.2820E+04 \(\pm\) 1.5606E+02

Relative difference using default slicing parameter: -7.7%
Factor of time the dipole implementation is faster: 3.4
\end{alltt}
\end{scriptsize}
\end{varwidth}
\end{center}
\caption{\label{tab:dipoleslicingcomp2}%
Same as in Table~\ref{tab:dipoleslicingcomp}, but for
$\sqrt{S}=7~\text{TeV}$, $6~\text{GeV}<p_T<8~\text{GeV}$, and $2.5<y<4$;
$\sqrt{S}=7~\text{TeV}$, $20~\text{GeV}<p_T<22~\text{GeV}$, and $-0.5<y<0.5$;
and
$\sqrt{S}=14~\text{TeV}$, $5~\text{GeV}<p_T<8~\text{GeV}$, and $2<y<4$.}
\end{table}

\begin{table}
\begin{center}
\begin{varwidth}{\linewidth}
\begin{scriptsize}
\begin{alltt}
p+\tol{p} --> c\tol{c}[3P21] + X:  Sqrt[s] = 100 GeV,  2 GeV < p_T < 3 GeV,  -0.5 < y < 0.5:
================================================================================

Dipole implementation:   3.8434E+02 \(\pm\) 3.8383E-01

Slicing implementation:
  0.000010    -1.9969E+04 \(\pm\) 6.1018E+00     2.0204E+04 \(\pm\) 8.2940E+00     2.3530E+02 \(\pm\) 1.0297E+01
  0.000032    -1.6852E+04 \(\pm\) 5.1391E+00     1.7156E+04 \(\pm\) 5.5671E+00     3.0380E+02 \(\pm\) 7.5765E+00
  0.000100    -1.4010E+04 \(\pm\) 4.0109E+00     1.4355E+04 \(\pm\) 3.6670E+00     3.4530E+02 \(\pm\) 5.4345E+00
  0.000320    -1.1317E+04 \(\pm\) 2.7969E+00     1.1679E+04 \(\pm\) 2.3758E+00     3.6180E+02 \(\pm\) 3.6697E+00
  0.001000    -8.8920E+03 \(\pm\) 2.0774E+00     9.2572E+03 \(\pm\) 1.5460E+00     3.6520E+02 \(\pm\) 2.5895E+00
  0.003200    -6.6371E+03 \(\pm\) 1.4326E+00     7.0012E+03 \(\pm\) 9.6310E-01     3.6410E+02 \(\pm\) 1.7262E+00
  0.010000    -4.6511E+03 \(\pm\) 9.3505E-01     5.0079E+03 \(\pm\) 5.8438E-01     3.5683E+02 \(\pm\) 1.1026E+00

Relative difference using default slicing parameter: -5.0%
Factor of time the dipole implementation is faster: 137

p+\tol{p} --> c\tol{c}[3P21] + X:  Sqrt[s] = 100 GeV,  6 GeV < p_T < 7 GeV,  -0.5 < y < 0.5:
================================================================================

Dipole implementation:  -1.1636E+00 \(\pm\) 1.1635E-03

Slicing implementation:
  0.000010    -2.5001E+01 \(\pm\) 4.9224E-03     2.3696E+01 \(\pm\) 1.4019E-02    -1.3046E+00 \(\pm\) 1.4858E-02
  0.000032    -2.1484E+01 \(\pm\) 3.6945E-03     2.0258E+01 \(\pm\) 9.5508E-03    -1.2264E+00 \(\pm\) 1.0240E-02
  0.000100    -1.8219E+01 \(\pm\) 3.1399E-03     1.7030E+01 \(\pm\) 6.2299E-03    -1.1887E+00 \(\pm\) 6.9764E-03
  0.000320    -1.5062E+01 \(\pm\) 2.4189E-03     1.3884E+01 \(\pm\) 4.1048E-03    -1.1778E+00 \(\pm\) 4.7645E-03
  0.001000    -1.2145E+01 \(\pm\) 1.8113E-03     1.0976E+01 \(\pm\) 2.7625E-03    -1.1687E+00 \(\pm\) 3.3033E-03
  0.003200    -9.3531E+00 \(\pm\) 1.3760E-03     8.2043E+00 \(\pm\) 1.6613E-03    -1.1488E+00 \(\pm\) 2.1572E-03
  0.010000    -6.8082E+00 \(\pm\) 9.9948E-04     5.7041E+00 \(\pm\) 1.0307E-03    -1.1041E+00 \(\pm\) 1.4357E-03

Relative difference using default slicing parameter: 0.4%
Factor of time the dipole implementation is faster: 4.3

p+\tol{p} --> c\tol{c}[3P21] + X:  Sqrt[s] = 1960 GeV,  7 GeV < p_T < 8 GeV,  -0.6 < y < 0.6:
=================================================================================

Dipole implementation:  -8.1975E+01 \(\pm\) 1.0328E-01

Slicing implementation:
  0.000010    -2.4577E+03 \(\pm\) 4.8300E-01     2.3627E+03 \(\pm\) 1.5549E+00    -9.4960E+01 \(\pm\) 1.6282E+00
  0.000032    -2.1184E+03 \(\pm\) 3.8233E-01     2.0317E+03 \(\pm\) 1.0440E+00    -8.6660E+01 \(\pm\) 1.1118E+00
  0.000100    -1.8032E+03 \(\pm\) 3.1323E-01     1.7206E+03 \(\pm\) 7.0188E-01    -8.2550E+01 \(\pm\) 7.6860E-01
  0.000320    -1.4983E+03 \(\pm\) 2.3867E-01     1.4167E+03 \(\pm\) 4.6731E-01    -8.1640E+01 \(\pm\) 5.2473E-01
  0.001000    -1.2168E+03 \(\pm\) 1.9288E-01     1.1358E+03 \(\pm\) 3.2025E-01    -8.1050E+01 \(\pm\) 3.7385E-01
  0.003200    -9.4671E+02 \(\pm\) 1.4560E-01     8.6743E+02 \(\pm\) 2.1682E-01    -7.9280E+01 \(\pm\) 2.6117E-01
  0.010000    -6.9965E+02 \(\pm\) 1.0889E-01     6.2487E+02 \(\pm\) 1.4362E-01    -7.4780E+01 \(\pm\) 1.8023E-01

Relative difference using default slicing parameter: -1.1%
Factor of time the dipole implementation is faster: 2.3
\end{alltt}
\end{scriptsize}
\end{varwidth}
\end{center}
\caption{\label{tab:dipoleslicingcomp3P21}%
Same as in Table~\ref{tab:dipoleslicingcomp}, but for
$\sigma(p\bar{p}\to c\bar{c}[{}^3\!P_2^{[1]}]+X)$.}
\end{table}

\begin{table}
\begin{center}
\begin{varwidth}{\linewidth}
\begin{scriptsize}
\begin{alltt}
p+\tol{p} --> c\tol{c}[3P21] + X:  Sqrt[s] = 7000 GeV,  6 GeV < p_T < 8 GeV,  2.5 < y < 4:
==============================================================================

Dipole implementation:  -5.0338E+02 \(\pm\) 8.1114E-01

Slicing implementation:
  0.000010    -2.1012E+04 \(\pm\) 4.3885E+00     2.0323E+04 \(\pm\) 1.4953E+01    -6.8870E+02 \(\pm\) 1.5584E+01
  0.000032    -1.8013E+04 \(\pm\) 3.3649E+00     1.7425E+04 \(\pm\) 9.6143E+00    -5.8830E+02 \(\pm\) 1.0186E+01
  0.000100    -1.5239E+04 \(\pm\) 2.8673E+00     1.4677E+04 \(\pm\) 6.5227E+00    -5.6230E+02 \(\pm\) 7.1251E+00
  0.000320    -1.2574E+04 \(\pm\) 2.2069E+00     1.2041E+04 \(\pm\) 4.2413E+00    -5.3310E+02 \(\pm\) 4.7811E+00
  0.001000    -1.0127E+04 \(\pm\) 1.6732E+00     9.6082E+03 \(\pm\) 2.8694E+00    -5.1900E+02 \(\pm\) 3.3216E+00
  0.003200    -7.7997E+03 \(\pm\) 1.2484E+00     7.2897E+03 \(\pm\) 1.9169E+00    -5.1000E+02 \(\pm\) 2.2876E+00
  0.010000    -5.6897E+03 \(\pm\) 8.9294E-01     5.2073E+03 \(\pm\) 1.2364E+00    -4.8240E+02 \(\pm\) 1.5251E+00

Relative difference using default slicing parameter: 3.1%
Factor of time the dipole implementation is faster: 2.1

p+\tol{p} --> c\tol{c}[3P21] + X:  Sqrt[s] = 7000 GeV,  20 GeV < p_T < 22 GeV,  -0.5 < y < 0.5:
===================================================================================

Dipole implementation:  -1.0014E+01 \(\pm\) 1.0013E-02

Slicing implementation:
  0.000010    -6.9249E+01 \(\pm\) 1.5399E-02     5.8884E+01 \(\pm\) 2.9578E-02    -1.0365E+01 \(\pm\) 3.3346E-02
  0.000032    -6.1220E+01 \(\pm\) 1.3769E-02     5.1169E+01 \(\pm\) 1.4606E-02    -1.0051E+01 \(\pm\) 2.0073E-02
  0.000100    -5.3471E+01 \(\pm\) 1.2195E-02     4.3475E+01 \(\pm\) 9.2384E-03    -9.9958E+00 \(\pm\) 1.5299E-02
  0.000320    -4.5673E+01 \(\pm\) 1.0564E-02     3.5722E+01 \(\pm\) 6.4647E-03    -9.9510E+00 \(\pm\) 1.2385E-02
  0.001000    -3.8146E+01 \(\pm\) 8.9563E-03     2.8256E+01 \(\pm\) 4.7210E-03    -9.8900E+00 \(\pm\) 1.0124E-02
  0.003200    -3.0578E+01 \(\pm\) 7.2845E-03     2.0851E+01 \(\pm\) 3.4411E-03    -9.7273E+00 \(\pm\) 8.0564E-03
  0.010000    -2.3281E+01 \(\pm\) 5.6306E-03     1.3958E+01 \(\pm\) 2.3551E-03    -9.3230E+00 \(\pm\) 6.1032E-03

Relative difference using default slicing parameter: -1.2%
Factor of time the dipole implementation is faster: 0.48

p+\tol{p} --> c\tol{c}[3P21] + X:  Sqrt[s] = 14000 GeV,  5 GeV < p_T < 8 GeV,  2 < y < 4:
=============================================================================

Dipole implementation:  -1.5363E+03 \(\pm\) 5.4212E+00

Slicing implementation:
  0.000010    -1.6753E+05 \(\pm\) 4.0603E+01     1.6446E+05 \(\pm\) 4.5001E+01    -3.0702E+03 \(\pm\) 6.0611E+01
  0.000032    -1.4322E+05 \(\pm\) 3.0717E+01     1.4094E+05 \(\pm\) 3.8639E+01    -2.2820E+03 \(\pm\) 4.9361E+01
  0.000100    -1.2083E+05 \(\pm\) 2.4876E+01     1.1889E+05 \(\pm\) 2.0610E+01    -1.9426E+03 \(\pm\) 3.2305E+01
  0.000320    -9.9425E+04 \(\pm\) 2.0367E+01     9.7638E+04 \(\pm\) 1.3883E+01    -1.7870E+03 \(\pm\) 2.4648E+01
  0.001000    -7.9823E+04 \(\pm\) 1.4390E+01     7.8121E+04 \(\pm\) 9.4865E+00    -1.7020E+03 \(\pm\) 1.7235E+01
  0.003200    -6.1268E+04 \(\pm\) 1.0435E+01     5.9614E+04 \(\pm\) 6.3932E+00    -1.6540E+03 \(\pm\) 1.2238E+01
  0.010000    -4.4537E+04 \(\pm\) 7.2279E+00     4.2986E+04 \(\pm\) 4.3827E+00    -1.5510E+03 \(\pm\) 8.4528E+00

Relative difference using default slicing parameter: 10.8%
Factor of time the dipole implementation is faster: 3.3
\end{alltt}
\end{scriptsize}
\end{varwidth}
\end{center}
\caption{\label{tab:dipoleslicingcomp23P21}%
Same as in Table~\ref{tab:dipoleslicingcomp2}, but for
$\sigma(p\bar{p}\to c\bar{c}[{}^3\!P_2^{[1]}]+X)$.}
\end{table}

\section{Summary\label{sec:summary}}

In this article, we have reviewed the singularity structure of NLO NRQCD calculations of the production of heavy-quark pairs in $S$ and $P$ wave states and
provided details of our phase space slicing implementation thereof.
Thereby, we have identified a common mistake in the literature.
Furthermore, we have summarized the dipole subtraction formalism for such calculations, which we have recently developed in Ref.~\cite{Butenschoen:2019lef}, added details about its implementation in terms of computer codes, and performed internal numeric tests.
Finally, we have extensively compared our two implementations numerically and found reasonable agreement.
As expected,  the dipole subtraction implementation outperforms the phase space slicing implementation both with regard to accuracy and speed.

\section*{Acknowledgments}

We would like to thank Anatoly Kotikov for assistance in evaluating a particular Feynman diagram.
This work was supported in part by the German Federal Ministry for Education
and Research BMBF through Grant No.\ 05H15GUCC1 and by the German Research
Foundation DFG through Grant No.\ KN~365/12-1.

\end{document}

%% file: TableOfDipoles.tex
\newcommand{\TODcolwidth}{119pt}
\newcommand{\TODscalefactor}{0.6}
\newcommand{\TODhspace}{0pt}
\begin{sidewaysfigure}
\begin{center}
\begin{minipage}[t]{\TODcolwidth}
\vspace{0pt}
\scalebox{\TODscalefactor}{
\begin{tabular}{cc}
\multicolumn{2}{c}{\bf $g+g\to c\overline{c}[n] + g:$} \\
\\
\multirow{2}{*}{
\begin{picture}(77,44)(0,0)
\Line(36,30)(72,30)
\Line(36,28)(72,28)
\Gluon(0,29)(36,29){1.5}{12}
\Gluon(0,15)(36,15){1.5}{12}
\Gluon(36,15)(72,15){1.5}{12}
\Gluon(56,15)(72,6){1.5}{6}
\GOval(36,22)(11,11)(0){0.9}
\Text(74,15)[l]{\scriptsize 3}
\Text(74,6)[l]{\scriptsize 4}
\end{picture}
}
& 1: $V_{g_3g_4,p_0}^{\mathrm{fin,}S_1}$ \\
& 2: $V_{g_3g_4,p_1}^{\mathrm{fin,}S_1}$ \\
& 3: $V_{g_3g_4,p_2}^{\mathrm{fin,}S_1}$ \\
\\
\multirow{2}{*}{
\begin{picture}(77,44)(0,0)
\Line(36,30)(72,30)
\Line(36,28)(72,28)
\Gluon(0,29)(36,29){1.5}{12}
\Gluon(0,15)(36,15){1.5}{12}
\Gluon(36,15)(57,15){1.5}{6}
\ArrowLine(57,15)(72,15)
\ArrowLine(72,6)(57,15)
\GOval(36,22)(11,11)(0){0.9}
\Text(74,15)[l]{\scriptsize 3}
\Text(74,6)[l]{\scriptsize 4}
\end{picture}
}
& 4: $V_{q_3\overline{q}_4,p_0}^{\mathrm{fin,}S_1}$ \\
& 5: $V_{q_3\overline{q}_4,p_1}^{\mathrm{fin,}S_1}$ \\
& 6: $V_{q_3\overline{q}_4,p_2}^{\mathrm{fin,}S_1}$ \\
\\
\multirow{2}{*}{
\begin{picture}(77,44)(0,0)
\Line(36,30)(72,30)
\Line(36,28)(72,28)
\Gluon(0,29)(36,29){1.5}{12}
\Gluon(0,15)(36,15){1.5}{12}
\Gluon(36,15)(72,15){1.5}{12}
\Gluon(57,29)(72,38){1.5}{6}
\GOval(36,22)(11,11)(0){0.9}
\Text(74,38)[l]{\scriptsize 3}
\Text(74,15)[l]{\scriptsize 4}
\end{picture}
}
& 7: $V_{p_0g_3,p_4}^{\mathrm{fin,}S_1}$ \\
& 8: $V_{p_0g_3,p_1}^{\mathrm{fin,}S_1}$ \\
& 9: $V_{p_0g_3,p_2}^{\mathrm{fin,}S_1}$ \\
\\
\multirow{2}{*}{
\begin{picture}(77,44)(0,0)
\Line(36,30)(72,30)
\Line(36,28)(72,28)
\Gluon(0,29)(36,29){1.5}{12}
\Gluon(0,15)(36,15){1.5}{12}
\Gluon(36,15)(72,15){1.5}{12}
\Gluon(57,29)(72,38){1.5}{6}
\GOval(36,22)(11,11)(0){0.9}
\Text(74,38)[l]{\scriptsize 4}
\Text(74,15)[l]{\scriptsize 3}
\end{picture}
}
& 10: $V_{p_0g_4,p_3}^{\mathrm{fin,}S_1}$ \\
& 11: $V_{p_0g_4,p_1}^{\mathrm{fin,}S_1}$ \\
& 12: $V_{p_0g_4,p_2}^{\mathrm{fin,}S_1}$ \\
\\
\multirow{2}{*}{
\begin{picture}(77,44)(0,0)
\Line(36,30)(72,30)
\Line(36,28)(72,28)
\Gluon(0,29)(36,29){1.5}{12}
\Gluon(0,15)(36,15){1.5}{12}
\Gluon(36,15)(72,15){1.5}{12}
\Gluon(20,38)(72,38){1.5}{18}
\GlueArc(23,29)(9,90,180){1.5}{4}
\GOval(36,22)(11,11)(0){0.9}
\Text(74,38)[l]{\scriptsize 3}
\Text(74,15)[l]{\scriptsize 4}
\end{picture}
}
& 13: $V_{g_1g_3,p_0}^{\mathrm{ini,}S_1}$ \\
& 14: $V_{g_1g_3,p_4}^{\mathrm{ini,}S_1}$ \\
& 15: $V_{g_1g_3,p_2}^{\mathrm{ini,}S_1}$ \\
\\
\multirow{2}{*}{
\begin{picture}(77,44)(0,0)
\Line(36,30)(72,30)
\Line(36,28)(72,28)
\Gluon(0,29)(36,29){1.5}{12}
\Gluon(0,15)(36,15){1.5}{12}
\Gluon(36,15)(72,15){1.5}{12}
\Gluon(20,38)(72,38){1.5}{18}
\GlueArc(23,29)(9,90,180){1.5}{4}
\GOval(36,22)(11,11)(0){0.9}
\Text(74,38)[l]{\scriptsize 4}
\Text(74,15)[l]{\scriptsize 3}
\end{picture}
}
& 16: $V_{g_1g_4,p_0}^{\mathrm{ini,}S_1}$ \\
& 17: $V_{g_1g_4,p_3}^{\mathrm{ini,}S_1}$ \\
& 18: $V_{g_1g_4,p_2}^{\mathrm{ini,}S_1}$ \\
\\
\multirow{2}{*}{
\begin{picture}(77,44)(0,0)
\Line(36,30)(72,30)
\Line(36,28)(72,28)
\Gluon(14,29)(36,29){1.5}{7}
\Gluon(0,15)(36,15){1.5}{12}
\Gluon(36,15)(72,15){1.5}{12}
\ArrowLine(23,38)(72,38)
\CArc(23,29)(9,90,180)
\ArrowLine(0,29)(14,29)
\GOval(36,22)(11,11)(0){0.9}
\Text(74,38)[l]{\scriptsize 3}
\Text(74,15)[l]{\scriptsize 4}
\end{picture}
}
& 19: $V_{q_1q_3,p_0}^{\mathrm{ini,}S_1}$ \\
& 20: $V_{q_1q_3,p_4}^{\mathrm{ini,}S_1}$ \\
& 21: $V_{q_1q_3,p_2}^{\mathrm{ini,}S_1}$ \\
\\
\multirow{2}{*}{
\begin{picture}(77,44)(0,0)
\Line(36,30)(72,30)
\Line(36,28)(72,28)
\Gluon(0,29)(36,29){1.5}{12}
\Gluon(0,15)(36,15){1.5}{12}
\Gluon(36,15)(72,15){1.5}{12}
\Gluon(20,6)(72,6){1.5}{18}
\GlueArc(23,15)(9,180,270){-1.5}{4}
\GOval(36,22)(11,11)(0){0.9}
\Text(74,15)[l]{\scriptsize 3}
\Text(74,6)[l]{\scriptsize 4}
\end{picture}
}
& 22: $V_{g_2g_4,p_0}^{\mathrm{ini,}S_1}$ \\
& 23: $V_{g_2g_4,p_3}^{\mathrm{ini,}S_1}$ \\
& 24: $V_{g_2g_4,p_1}^{\mathrm{ini,}S_1}$ \\
\\
\multirow{2}{*}{
\begin{picture}(77,44)(0,0)
\Line(36,30)(72,30)
\Line(36,28)(72,28)
\Gluon(0,29)(36,29){1.5}{12}
\Gluon(0,15)(36,15){1.5}{12}
\Gluon(36,15)(72,15){1.5}{12}
\Gluon(20,6)(72,6){1.5}{18}
\GlueArc(23,15)(9,180,270){-1.5}{4}
\GOval(36,22)(11,11)(0){0.9}
\Text(74,15)[l]{\scriptsize 4}
\Text(74,6)[l]{\scriptsize 3}
\end{picture}
}
& 25: $V_{g_2g_3,p_0}^{\mathrm{ini,}S_1}$ \\
& 26: $V_{g_2g_3,p_4}^{\mathrm{ini,}S_1}$ \\
& 27: $V_{g_2g_3,p_1}^{\mathrm{ini,}S_1}$ \\
\\
\multirow{2}{*}{
\begin{picture}(77,44)(0,0)
\Line(36,30)(72,30)
\Line(36,28)(72,28)
\Gluon(0,29)(36,29){1.5}{12}
\Gluon(14,15)(36,15){1.5}{7}
\Gluon(36,15)(72,15){1.5}{12}
\ArrowLine(0,15)(14,15)
\CArc(23,15)(9,180,270)
\ArrowLine(23,6)(72,6)
\GOval(36,22)(11,11)(0){0.9}
\Text(74,15)[l]{\scriptsize 4}
\Text(74,6)[l]{\scriptsize 3}
\end{picture}
}
& 28: $V_{q_2q_3,p_0}^{\mathrm{ini,}S_1}$ \\
& 29: $V_{q_2q_3,p_4}^{\mathrm{ini,}S_1}$ \\
& 30: $V_{q_2q_3,p_1}^{\mathrm{ini,}S_1}$
\end{tabular}
}
\end{minipage}
\hspace{\TODhspace}
\begin{minipage}[t]{\TODcolwidth}
\vspace{0pt}
\scalebox{\TODscalefactor}{
\begin{tabular}{cc}
\multicolumn{2}{c}{\bf $q+\overline{q}\to c\overline{c}[n] + g:$} \\
\\
\multirow{2}{*}{
\begin{picture}(77,44)(0,0)
\Line(36,30)(72,30)
\Line(36,28)(72,28)
\ArrowLine(0,29)(28,29)
\ArrowLine(28,15)(0,15)
\Gluon(36,15)(72,15){1.5}{12}
\Gluon(56,15)(72,6){1.5}{6}
\GOval(36,22)(11,11)(0){0.9}
\Text(74,15)[l]{\scriptsize 3}
\Text(74,6)[l]{\scriptsize 4}
\end{picture}
}
& 1: $V_{g_3g_4,p_0}^{\mathrm{fin,}S_1}$ \\
& 2: $V_{g_3g_4,p_1}^{\mathrm{fin,}S_1}$ \\
& 3: $V_{g_3g_4,p_2}^{\mathrm{fin,}S_1}$ \\
\\
\multirow{2}{*}{
\begin{picture}(77,44)(0,0)
\Line(36,30)(72,30)
\Line(36,28)(72,28)
\ArrowLine(0,29)(28,29)
\ArrowLine(28,15)(0,15)
\Gluon(36,15)(57,15){1.5}{6}
\ArrowLine(57,15)(72,15)
\ArrowLine(72,6)(57,15)
\GOval(36,22)(11,11)(0){0.9}
\Text(74,15)[l]{\scriptsize 3}
\Text(74,6)[l]{\scriptsize 4}
\end{picture}
}
& 4: $V_{q_3\overline{q}_4,p_0}^{\mathrm{fin,}S_1}$ \\
& 5: $V_{q_3\overline{q}_4,p_1}^{\mathrm{fin,}S_1}$ \\
& 6: $V_{q_3\overline{q}_4,p_2}^{\mathrm{fin,}S_1}$ \\
\\
\multirow{2}{*}{
\begin{picture}(77,44)(0,0)
\Line(36,30)(72,30)
\Line(36,28)(72,28)
\ArrowLine(0,29)(28,29)
\ArrowLine(28,15)(0,15)
\Gluon(36,15)(72,15){1.5}{12}
\Gluon(57,29)(72,38){1.5}{6}
\GOval(36,22)(11,11)(0){0.9}
\Text(74,38)[l]{\scriptsize 3}
\Text(74,15)[l]{\scriptsize 4}
\end{picture}
}
& 7: $V_{p_0g_3,p_4}^{\mathrm{fin,}S_1}$ \\
& 8: $V_{p_0g_3,p_1}^{\mathrm{fin,}S_1}$ \\
& 9: $V_{p_0g_3,p_2}^{\mathrm{fin,}S_1}$ \\
\\
\multirow{2}{*}{
\begin{picture}(77,44)(0,0)
\Line(36,30)(72,30)
\Line(36,28)(72,28)
\ArrowLine(0,29)(28,29)
\ArrowLine(28,15)(0,15)
\Gluon(36,15)(72,15){1.5}{12}
\Gluon(57,29)(72,38){1.5}{6}
\GOval(36,22)(11,11)(0){0.9}
\Text(74,38)[l]{\scriptsize 4}
\Text(74,15)[l]{\scriptsize 3}
\end{picture}
}
& 10: $V_{p_0g_4,p_3}^{\mathrm{fin,}S_1}$ \\
& 11: $V_{p_0g_4,p_1}^{\mathrm{fin,}S_1}$ \\
& 12: $V_{p_0g_4,p_2}^{\mathrm{fin,}S_1}$ \\
\\
\multirow{2}{*}{
\begin{picture}(77,44)(0,0)
\Line(36,30)(72,30)
\Line(36,28)(72,28)
\ArrowLine(0,29)(14,29)
\ArrowLine(14,29)(28,29)
\ArrowLine(28,15)(0,15)
\Gluon(36,15)(72,15){1.5}{12}
\Gluon(20,38)(72,38){1.5}{18}
\GlueArc(23,29)(9,90,180){1.5}{4}
\GOval(36,22)(11,11)(0){0.9}
\Text(74,38)[l]{\scriptsize 3}
\Text(74,15)[l]{\scriptsize 4}
\end{picture}
}
& 13: $V_{q_1g_3,p_0}^{\mathrm{ini,}S_1}$ \\
& 14: $V_{q_1g_3,p_4}^{\mathrm{ini,}S_1}$ \\
& 15: $V_{q_1g_3,p_2}^{\mathrm{ini,}S_1}$ \\
\\
\multirow{2}{*}{
\begin{picture}(77,44)(0,0)
\Line(36,30)(72,30)
\Line(36,28)(72,28)
\ArrowLine(0,29)(14,29)
\ArrowLine(14,29)(28,29)
\ArrowLine(28,15)(0,15)
\Gluon(36,15)(72,15){1.5}{12}
\Gluon(20,38)(72,38){1.5}{18}
\GlueArc(23,29)(9,90,180){1.5}{4}
\GOval(36,22)(11,11)(0){0.9}
\Text(74,38)[l]{\scriptsize 4}
\Text(74,15)[l]{\scriptsize 3}
\end{picture}
}
& 16: $V_{q_1g_4,p_0}^{\mathrm{ini,}S_1}$ \\
& 17: $V_{q_1g_4,p_3}^{\mathrm{ini,}S_1}$ \\
& 18: $V_{q_1g_4,p_2}^{\mathrm{ini,}S_1}$ \\
\\
\multirow{2}{*}{
\begin{picture}(77,44)(0,0)
\Line(36,30)(72,30)
\Line(36,28)(72,28)
\ArrowLine(14,29)(28,29)
\Gluon(0,29)(14,29){1.5}{4}
\ArrowLine(28,15)(0,15)
\Gluon(36,15)(72,15){1.5}{12}
\ArrowLine(72,38)(23,38)
\CArc(23,29)(9,90,180)
\GOval(36,22)(11,11)(0){0.9}
\Text(74,38)[l]{\scriptsize 3}
\Text(74,15)[l]{\scriptsize 4}
\end{picture}
}
& 19: $V_{g_1\overline{q}_3,p_0}^{\mathrm{ini,}S_1}$ \\
& 20: $V_{g_1\overline{q}_3,p_4}^{\mathrm{ini,}S_1}$ \\
& 21: $V_{g_1\overline{q}_3,p_2}^{\mathrm{ini,}S_1}$ \\
\\
\multirow{2}{*}{
\begin{picture}(77,44)(0,0)
\Line(36,30)(72,30)
\Line(36,28)(72,28)
\ArrowLine(0,29)(28,29)
\ArrowLine(28,15)(14,15)
\ArrowLine(14,15)(0,15)
\Gluon(36,15)(72,15){1.5}{12}
\Gluon(20,6)(72,6){1.5}{18}
\GlueArc(23,15)(9,180,270){-1.5}{4}
\GOval(36,22)(11,11)(0){0.9}
\Text(74,15)[l]{\scriptsize 3}
\Text(74,6)[l]{\scriptsize 4}
\end{picture}
}
& 22: $V_{q_2g_4,p_0}^{\mathrm{ini,}S_1}$ \\
& 23: $V_{q_2g_4,p_3}^{\mathrm{ini,}S_1}$ \\
& 24: $V_{q_2g_4,p_1}^{\mathrm{ini,}S_1}$ \\
\\
\multirow{2}{*}{
\begin{picture}(77,44)(0,0)
\Line(36,30)(72,30)
\Line(36,28)(72,28)
\ArrowLine(0,29)(28,29)
\ArrowLine(28,15)(14,15)
\ArrowLine(14,15)(0,15)
\Gluon(36,15)(72,15){1.5}{12}
\Gluon(20,6)(72,6){1.5}{18}
\GlueArc(23,15)(9,180,270){-1.5}{4}
\GOval(36,22)(11,11)(0){0.9}
\Text(74,15)[l]{\scriptsize 4}
\Text(74,6)[l]{\scriptsize 3}
\end{picture}
}
& 25: $V_{q_2g_3,p_0}^{\mathrm{ini,}S_1}$ \\
& 26: $V_{q_2g_3,p_4}^{\mathrm{ini,}S_1}$ \\
& 27: $V_{q_2g_3,p_1}^{\mathrm{ini,}S_1}$ \\
\\
\multirow{2}{*}{
\begin{picture}(77,44)(0,0)
\Line(36,30)(72,30)
\Line(36,28)(72,28)
\ArrowLine(0,29)(28,29)
\ArrowLine(28,15)(14,15)
\Gluon(36,15)(72,15){1.5}{12}
\Gluon(0,15)(14,15){1.5}{4}
\CArc(23,15)(9,180,270)
\ArrowLine(23,6)(72,6)
\GOval(36,22)(11,11)(0){0.9}
\Text(74,15)[l]{\scriptsize 4}
\Text(74,6)[l]{\scriptsize 3}
\end{picture}
}
& 28: $V_{g_2\overline{q}_3,p_0}^{\mathrm{ini,}S_1}$ \\
& 29: $V_{g_2\overline{q}_3,p_4}^{\mathrm{ini,}S_1}$ \\
& 30: $V_{g_2\overline{q}_3,p_1}^{\mathrm{ini,}S_1}$
\end{tabular}
}
\end{minipage}
\hspace{\TODhspace}
\begin{minipage}[t]{\TODcolwidth}
\vspace{0pt}
\scalebox{\TODscalefactor}{
\begin{tabular}{cc}
\multicolumn{2}{c}{\bf $q+g\to c\overline{c}[n] + q:$} \\
\\
\multirow{2}{*}{
\begin{picture}(77,44)(0,0)
\Line(36,30)(72,30)
\Line(36,28)(72,28)
\ArrowLine(0,29)(28,29)
\Gluon(0,15)(36,15){1.5}{12}
\ArrowLine(42,15)(60,15)
\ArrowLine(60,15)(72,15)
\Gluon(56,15)(72,6){1.5}{6}
\GOval(36,22)(11,11)(0){0.9}
\Text(74,15)[l]{\scriptsize 3}
\Text(74,6)[l]{\scriptsize 4}
\end{picture}
}
& 1: $V_{q_3g_4,p_0}^{\mathrm{fin,}S_1}$ \\
& 2: $V_{q_3g_4,p_2}^{\mathrm{fin,}S_1}$ \\
& 3: $V_{q_3g_4,p_1}^{\mathrm{fin,}S_1}$ \\
\\
\multirow{2}{*}{
\begin{picture}(77,44)(0,0)
\Line(36,30)(72,30)
\Line(36,28)(72,28)
\ArrowLine(0,29)(28,29)
\Gluon(0,15)(36,15){1.5}{12}
\ArrowLine(42,15)(72,15)
\Gluon(57,29)(72,38){1.5}{6}
\GOval(36,22)(11,11)(0){0.9}
\Text(74,38)[l]{\scriptsize 4}
\Text(74,15)[l]{\scriptsize 3}
\end{picture}
}
& 4: $V_{p_0g_4,p_3}^{\mathrm{fin,}S_1}$ \\
& 5: $V_{p_0g_4,p_2}^{\mathrm{fin,}S_1}$ \\
& 6: $V_{p_0g_4,p_1}^{\mathrm{fin,}S_1}$ \\
\\
\multirow{2}{*}{
\begin{picture}(77,44)(0,0)
\Line(36,30)(72,30)
\Line(36,28)(72,28)
\ArrowLine(0,29)(14,29)
\ArrowLine(14,29)(28,29)
\Gluon(0,15)(36,15){1.5}{12}
\ArrowLine(42,15)(72,15)
\Gluon(20,38)(72,38){1.5}{18}
\GlueArc(23,29)(9,90,180){1.5}{4}
\GOval(36,22)(11,11)(0){0.9}
\Text(74,38)[l]{\scriptsize 4}
\Text(74,15)[l]{\scriptsize 3}
\end{picture}
}
& 7: $V_{q_1g_4,p_0}^{\mathrm{ini,}S_1}$ \\
& 8: $V_{q_1g_4,p_3}^{\mathrm{ini,}S_1}$ \\
& 9: $V_{q_1g_4,p_2}^{\mathrm{ini,}S_1}$ \\
\\
\multirow{2}{*}{
\begin{picture}(77,44)(0,0)
\Line(36,30)(72,30)
\Line(36,28)(72,28)
\Gluon(0,15)(36,15){1.5}{12}
\ArrowLine(42,15)(72,15)
\ArrowLine(72,38)(23,38)
\CArc(23,29)(9,90,180)
\ArrowLine(14,29)(28,29)
\Gluon(0,29)(14,29){1.5}{4}
\GOval(36,22)(11,11)(0){0.9}
\Text(74,38)[l]{\scriptsize 4}
\Text(74,15)[l]{\scriptsize 3}
\end{picture}
}
& 10: $V_{g_1\overline{q}_4,p_0}^{\mathrm{ini,}S_1}$ \\
& 11: $V_{g_1\overline{q}_4,p_3}^{\mathrm{ini,}S_1}$ \\
& 12: $V_{g_1\overline{q}_4,p_2}^{\mathrm{ini,}S_1}$ \\
\\
\multirow{2}{*}{
\begin{picture}(77,44)(0,0)
\Line(36,30)(72,30)
\Line(36,28)(72,28)
\ArrowLine(0,29)(28,29)
\Gluon(0,15)(36,15){1.5}{12}
\ArrowLine(42,15)(72,15)
\Gluon(20,6)(72,6){1.5}{18}
\GlueArc(23,15)(9,180,270){-1.5}{4}
\GOval(36,22)(11,11)(0){0.9}
\Text(74,15)[l]{\scriptsize 3}
\Text(74,6)[l]{\scriptsize 4}
\end{picture}
}
& 13: $V_{g_2g_4,p_0}^{\mathrm{ini,}S_1}$ \\
& 14: $V_{g_2g_4,p_3}^{\mathrm{ini,}S_1}$ \\
& 15: $V_{g_2g_4,p_1}^{\mathrm{ini,}S_1}$ \\
\\
\multirow{2}{*}{
\begin{picture}(77,44)(0,0)
\Line(36,30)(72,30)
\Line(36,28)(72,28)
\ArrowLine(0,29)(28,29)
\Gluon(14,15)(36,15){1.5}{7}
\ArrowLine(42,15)(72,15)
\ArrowLine(0,15)(14,15)
\CArc(23,15)(9,180,270)
\ArrowLine(23,6)(72,6)
\GOval(36,22)(11,11)(0){0.9}
\Text(74,15)[l]{\scriptsize 3}
\Text(74,6)[l]{\scriptsize 4}
\end{picture}
}
& 16: $V_{q_2q_4,p_0}^{\mathrm{ini,}S_1}$ \\
& 17: $V_{q_2q_4,p_3}^{\mathrm{ini,}S_1}$ \\
& 18: $V_{q_2q_4,p_1}^{\mathrm{ini,}S_1}$ \\
\\
\multirow{2}{*}{
\begin{picture}(77,44)(0,0)
\Line(36,30)(72,30)
\Line(36,28)(72,28)
\ArrowLine(0,29)(28,29)
\Gluon(14,15)(36,15){1.5}{7}
\ArrowLine(42,15)(72,15)
\ArrowLine(0,15)(14,15)
\CArc(23,15)(9,180,270)
\ArrowLine(23,6)(72,6)
\GOval(36,22)(11,11)(0){0.9}
\Text(74,15)[l]{\scriptsize 4}
\Text(74,6)[l]{\scriptsize 3}
\end{picture}
}
& 19: $V_{q_2q_3,p_0}^{\mathrm{ini,}S_1}$ \\
& 20: $V_{q_2q_3,p_4}^{\mathrm{ini,}S_1}$ \\
& 21: $V_{q_2q_3,p_1}^{\mathrm{ini,}S_1}$ \\
\multicolumn{2}{c}{\small(equal quark flavors only)}
\end{tabular}
}
\end{minipage}
\hspace{\TODhspace}
\begin{minipage}[t]{\TODcolwidth}
\vspace{0pt}
\scalebox{\TODscalefactor}{
\begin{tabular}{cc}
\multicolumn{2}{c}{\bf $g+q\to c\overline{c}[n] + q:$} \\
\\
\multirow{2}{*}{
\begin{picture}(77,44)(0,0)
\Line(36,30)(72,30)
\Line(36,28)(72,28)
\ArrowLine(0,15)(28,15)
\Gluon(0,29)(36,29){1.5}{12}
\ArrowLine(42,15)(60,15)
\ArrowLine(60,15)(72,15)
\Gluon(56,15)(72,6){1.5}{6}
\GOval(36,22)(11,11)(0){0.9}
\Text(74,15)[l]{\scriptsize 3}
\Text(74,6)[l]{\scriptsize 4}
\end{picture}
}
& 1: $V_{q_3g_4,p_0}^{\mathrm{fin,}S_1}$ \\
& 2: $V_{q_3g_4,p_2}^{\mathrm{fin,}S_1}$ \\
& 3: $V_{q_3g_4,p_1}^{\mathrm{fin,}S_1}$ \\
\\
\multirow{2}{*}{
\begin{picture}(77,44)(0,0)
\Line(36,30)(72,30)
\Line(36,28)(72,28)
\ArrowLine(0,15)(28,15)
\Gluon(0,29)(36,29){1.5}{12}
\ArrowLine(42,15)(72,15)
\Gluon(57,29)(72,38){1.5}{6}
\GOval(36,22)(11,11)(0){0.9}
\Text(74,38)[l]{\scriptsize 4}
\Text(74,15)[l]{\scriptsize 3}
\end{picture}
}
& 4: $V_{p_0g_4,p_3}^{\mathrm{fin,}S_1}$ \\
& 5: $V_{p_0g_4,p_2}^{\mathrm{fin,}S_1}$ \\
& 6: $V_{p_0g_4,p_1}^{\mathrm{fin,}S_1}$ \\
\\
\multirow{2}{*}{
\begin{picture}(77,44)(0,0)
\Line(36,30)(72,30)
\Line(36,28)(72,28)
\Gluon(0,29)(36,29){1.5}{12}
\ArrowLine(0,15)(28,15)
\ArrowLine(42,15)(72,15)
\Gluon(20,38)(72,38){1.5}{18}
\GlueArc(23,29)(9,90,180){1.5}{4}
\GOval(36,22)(11,11)(0){0.9}
\Text(74,38)[l]{\scriptsize 4}
\Text(74,15)[l]{\scriptsize 3}
\end{picture}
}
& 7: $V_{g_1g_4,p_0}^{\mathrm{ini,}S_1}$ \\
& 8: $V_{g_1g_4,p_3}^{\mathrm{ini,}S_1}$ \\
& 9: $V_{g_1g_4,p_2}^{\mathrm{ini,}S_1}$ \\
\\
\multirow{2}{*}{
\begin{picture}(77,44)(0,0)
\Line(36,30)(72,30)
\Line(36,28)(72,28)
\Gluon(14,29)(36,29){1.5}{7}
\ArrowLine(0,15)(28,15)
\ArrowLine(42,15)(72,15)
\ArrowLine(23,38)(72,38)
\CArc(23,29)(9,90,180)
\ArrowLine(0,29)(14,29)
\GOval(36,22)(11,11)(0){0.9}
\Text(74,38)[l]{\scriptsize 4}
\Text(74,15)[l]{\scriptsize 3}
\end{picture}
}
& 10: $V_{q_1q_4,p_0}^{\mathrm{ini,}S_1}$ \\
& 11: $V_{q_1q_4,p_3}^{\mathrm{ini,}S_1}$ \\
& 12: $V_{q_1q_4,p_2}^{\mathrm{ini,}S_1}$ \\
\multicolumn{2}{c}{\small(equal quark flavors only)} \\
\\
\multirow{2}{*}{
\begin{picture}(77,44)(0,0)
\Line(36,30)(72,30)
\Line(36,28)(72,28)
\Gluon(14,29)(36,29){1.5}{7}
\ArrowLine(0,15)(28,15)
\ArrowLine(42,15)(72,15)
\ArrowLine(23,38)(72,38)
\CArc(23,29)(9,90,180)
\ArrowLine(0,29)(14,29)
\GOval(36,22)(11,11)(0){0.9}
\Text(74,38)[l]{\scriptsize 3}
\Text(74,15)[l]{\scriptsize 4}
\end{picture}
}
& 13: $V_{q_1q_3,p_0}^{\mathrm{ini,}S_1}$ \\
& 14: $V_{q_1q_3,p_4}^{\mathrm{ini,}S_1}$ \\
& 15: $V_{q_1q_3,p_2}^{\mathrm{ini,}S_1}$ \\
\\
\multirow{2}{*}{
\begin{picture}(77,44)(0,0)
\Line(36,30)(72,30)
\Line(36,28)(72,28)
\ArrowLine(0,15)(14,15)
\ArrowLine(14,15)(28,15)
\Gluon(0,29)(36,29){1.5}{12}
\ArrowLine(42,15)(72,15)
\Gluon(20,6)(72,6){1.5}{18}
\GlueArc(23,15)(9,180,270){-1.5}{4}
\GOval(36,22)(11,11)(0){0.9}
\Text(74,15)[l]{\scriptsize 3}
\Text(74,6)[l]{\scriptsize 4}
\end{picture}
}
& 16: $V_{q_2g_4,p_0}^{\mathrm{ini,}S_1}$ \\
& 17: $V_{q_2g_4,p_3}^{\mathrm{ini,}S_1}$ \\
& 18: $V_{q_2g_4,p_1}^{\mathrm{ini,}S_1}$ \\
\\
\multirow{2}{*}{
\begin{picture}(77,44)(0,0)
\Line(36,30)(72,30)
\Line(36,28)(72,28)
\Gluon(0,29)(36,29){1.5}{12}
\ArrowLine(14,15)(28,15)
\ArrowLine(42,15)(72,15)
\Gluon(0,15)(14,15){1.5}{4}
\CArc(23,15)(9,180,270)
\ArrowLine(72,6)(23,6)
\GOval(36,22)(11,11)(0){0.9}
\Text(74,15)[l]{\scriptsize 3}
\Text(74,6)[l]{\scriptsize 4}
\end{picture}
}
& 19: $V_{g_2\overline{q}_4,p_0}^{\mathrm{ini,}S_1}$ \\
& 20: $V_{g_2\overline{q}_4,p_3}^{\mathrm{ini,}S_1}$ \\
& 21: $V_{g_2\overline{q}_4,p_1}^{\mathrm{ini,}S_1}$ 
\end{tabular}
}
\end{minipage}
\hspace{\TODhspace}
\begin{minipage}[t]{\TODcolwidth}
\vspace{0pt}
\scalebox{\TODscalefactor}{
\begin{tabular}{c}
Additional dipoles for $P$-wave states\\
(include for all subprocesses): \\
\\
231: $V_{S_2,31}$ \\
232: $V_{S_2,32}$ \\
234: $V_{S_2,34}$ \\
\\
241: $V_{S_2,41}$ \\
242: $V_{S_2,42}$ \\
243: $V_{S_2,43}$ \\
\\
330: $V_{S_3,3}$ \\
340: $V_{S_3,4}$ \\
\end{tabular}
}
\end{minipage}
\end{center}
\caption{\label{fig:listofdipoles}
Numbered list of dipole terms for each of the occurring Born processes with $2\to 2$ kinematics. The diagrams related to the $V_{ij,k}^{\mathrm{ini,}S_1}$ and $V_{ij,k}^{\mathrm{fin,}S_1}$ terms indicate in which collinear or soft limits the latter contribute. Light-quark lines are to be summed over all quark flavors.}
\end{sidewaysfigure}
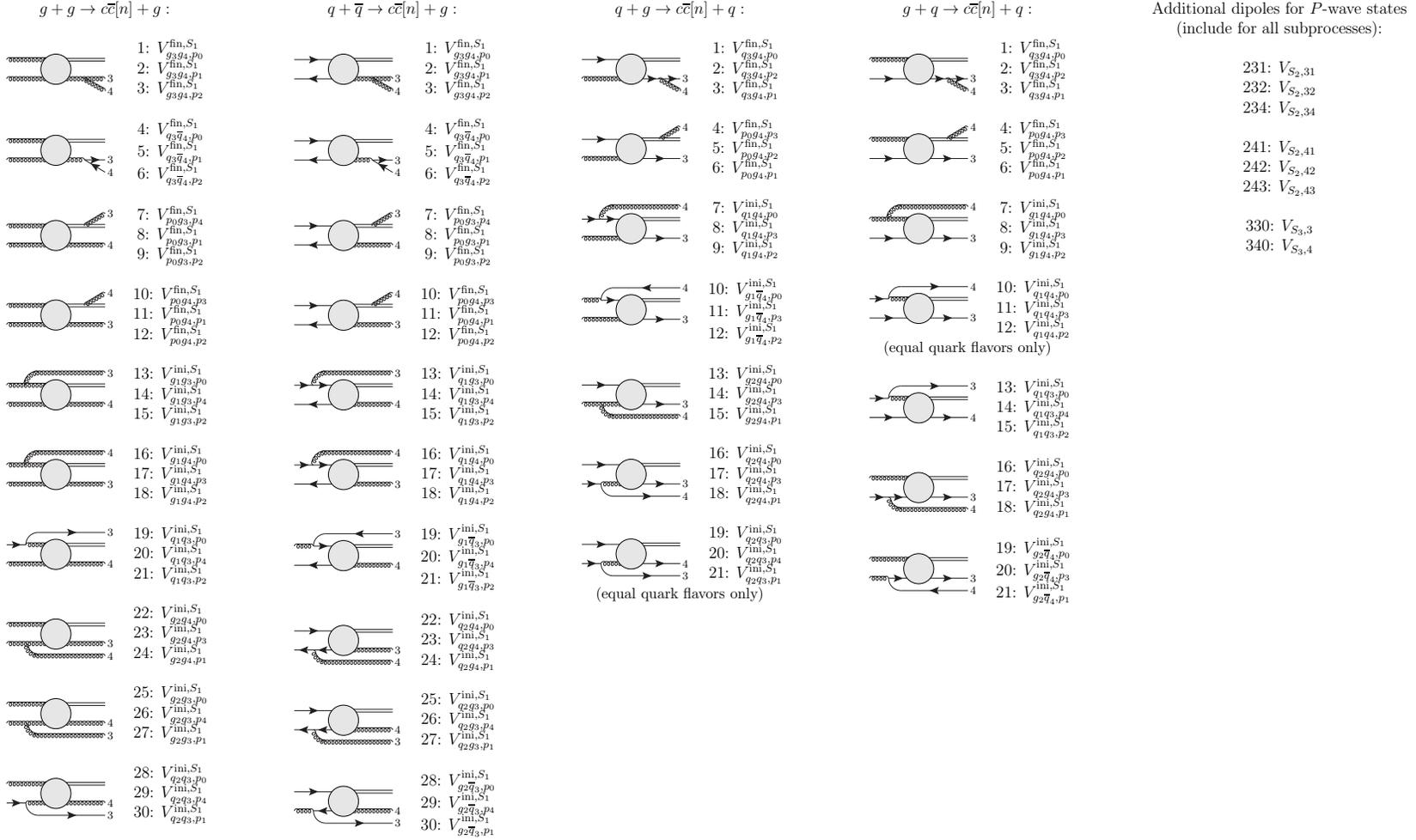